\documentclass[a4paper,fleqn,usenatbib]{mnras}

\usepackage{savesym}
\usepackage{amsmath}
\savesymbol{iint}
\savesymbol{iiint}
\usepackage{txfonts}
\restoresymbol{TXF}{iint}
\restoresymbol{TXF}{iiint}

\usepackage[T1]{fontenc}
\usepackage{ae,aecompl}

\usepackage{graphicx}
\usepackage{amsmath}
\usepackage{amssymb}	% Extra maths symbols
\bibpunct{(}{)}{;}{a}{}{,}
 
\title[The globular cluster system of NGC\,6411]{The globular cluster system of the isolated elliptical NGC\,6411: An old system in an intermediate-age galaxy? \thanks{Based on observations
obtained at the Gemini Observatory, which is operated by the 
Association of Universities for Research in Astronomy, Inc., under a 
cooperative agreement with the NSF on behalf of the Gemini partnership: the 
National Science Foundation (United States), the National Research Council 
(Canada), CONICYT (Chile), Ministerio de Ciencia, Tecnolog\'ia e Innovaci\'on 
Productiva (Argentina), and Ministerio da Ciencia, Tecnologia e Inovasao 
(Brazil).}}
\author[J. P. Caso et al.]{Juan Pablo Caso$^{~1,2}$\thanks{E-mails: jpcaso@fcaglp.unlp.edu.ar (JPC); lbassino@fcaglp.unlp.edu.ar (LB); tom@astro-udec.cl (TR); rsalinas@gemini.edu (RS)}, Lilia P. Bassino$^{~1,2}$, Tom Richtler$^{~3}$ and Ricardo Salinas$^{~4}$\\
$^{~1}$Facultad de Ciencias Astron\'omicas y Geof\'isicas de la Universidad Nacional de     
La Plata, and Instituto de Astrof\'isica de La Plata \\
(CCT La Plata -- CONICET, UNLP), Paseo del Bosque S/N, B1900FWA La Plata, Argentina\\   
$^{~2}$Consejo Nacional de Investigaciones Cient\'ificas y T\'ecnicas, Rivadavia 1917, 
C1033AAJ Ciudad Aut\'onoma de Buenos Aires,\\
Argentina\\   
$^{~3}$Departamento de Astronom\'ia, Universidad de Concepci\'on, Concepci\'on, Chile\\
$^{~4}$Gemini Observatory, Casilla 603, La Serena, Chile}

\date{Accepted XXX. Received YYY; in original form ZZZ}

\pubyear{2018}  

\begin{document}
\label{firstpage}
\pagerange{\pageref{firstpage}--\pageref{lastpage}}
\maketitle

\begin{abstract}
We present a photometric study of the isolated elliptical NGC\,6411 and 
its globular cluster (GC) system, based on Gemini/GMOS $g'$, $r'$, $i'$ 
photometry. This galaxy has been the host of a SNIa. Spectral studies 
indicate an intermediate-age and very metal-rich stellar population in 
its inner region. The extent of the globular cluster system is about 65 kpc. 
Its population contains $700\pm45$ members, implying a rather poor system, as 
normally found for isolated ellipticals. 
The colour bimodality and luminosity distribution are typical of old GC 
systems.  
An excess of bright GCs with intermediate colours might evidence an
intermediate-age merger, but their colours indicate an older event than 
ages derived from spectroscopic studies of the diffuse
light of the galaxy.
\end{abstract}

\begin{keywords}
galaxies: elliptical and lenticular, cD -- galaxies: evolution -- galaxies: star clusters: individual: NGC\,6411

\end{keywords}

\section{Introduction}
\label{intro}
Elliptical galaxies (Es) usually inhabit groups and clusters of galaxies
\citep{dre80,bam09}, where the density of neighbours favoured merging 
processes, which are thought to be the main source of mass accretion for 
bright ellipticals since $z\approx 2$ \citep[e.g.][]{vdo10,jim11,sch14}. 
On the other hand, isolated ellipticals (iEs) might have experienced less 
mergers, evolving more quietly. From numerical simulations \citet{hir13} 
indicated that iEs have been shaped by late mergers in larger proportion 
than Es in denser environments. Similar results have been reported by 
\citet{nie10}, who also found that iEs in the Millenium Simulation are 
typically bluer and present lighter haloes than Es in clusters.
These results are supported by 
observational studies who found tidal features in a large fraction of iEs 
\citep[e.g.][]{her08,tal09}, bluer colours than Es in clusters \citep{lac16} 
and evidence for light dark matter haloes \citep{men09,sal12,ric15}, but 
in some cases results differ, pointing to a variety of evolutionary histories 
\citep[e.g.][]{marc04}.

As the study of Es has been biased by high-density environments, the 
analyses of globular cluster systems (GCSs) in iEs have been limited 
to a few cases \citep[e.g.][]{spi08,lan13,ric15}. This is in part motivated 
by the small number of iEs found in the local Universe, but also due to the 
rather poor GCSs they normally have \citep[][and references therein]{cas13b,sal15} 
in comparison with bright Es in clusters of galaxies \citep{har15,cas17}. 

It is believed that the origin of globular clusters (GCs) is directly 
related to the merger history of the galaxy they belong to, either by the 
extreme conditions needed to form GCs \citep{kru14,kru15}, or by the
accretion of GCs during the merging process \citep{ton13}. Hence, the 
study of GCSs might bring evidence about the evolutionary history of a 
galaxy \citep[e.g.][]{cas15a,ric12a,ric13,ric14,bas17}.
In the present paper we perform a study of the iE NGC\,6411, a 
moderately bright galaxy with an estimated distance of 40\,Mpc 
based on surface-brightness fluctuations \citep[SBF][]{bla01}. It 
hosted a supernovae Ia \citep[SN 1999da][]{fil99} at 75\,arcsec 
from the galaxy centre, classified as a subluminous SNIa. 
The distances derived in this latter way are larger, ranging from 50 
to 70\,Mpc according to NED\footnote{This research has made use of the 
NASA/IPAC Extragalactic Database (NED) which is operated by the Jet 
Propulsion Laboratory, California Institute of Technology, under 
contract with the National Aeronautics and Space Administration.}.
Spectroscopic analysis from the CALIFA survey \citep{gon15} 
reveals an intermediate age of  $\approx 3.5$\,Gyr and a high 
metallicity of $\log_{10}(Z/Z_{\odot})=0.28$, if fitted as a single stellar 
population. 
\citet{san06} also analysed spectroscopic observations from NGC\,6411, 
obtaining a similar metallicity for the central region, but a
slightly older population, $\approx 5$\,Gyr. 
Our aim is to characterise the GCS of the
galaxy, in order to contribute to the analysis of its evolutionary history.

This paper is organised as follows. The observations and data
reduction are described in Section\,2, the results are presented in
Section\,3 and Section\,4 is devoted to the discussion. Finally, a
summary and the concluding remarks are given in Section\,5.

\section[]{Observations and data reduction}

\subsection{Observations}
\begin{figure}
 \includegraphics[width=80mm]{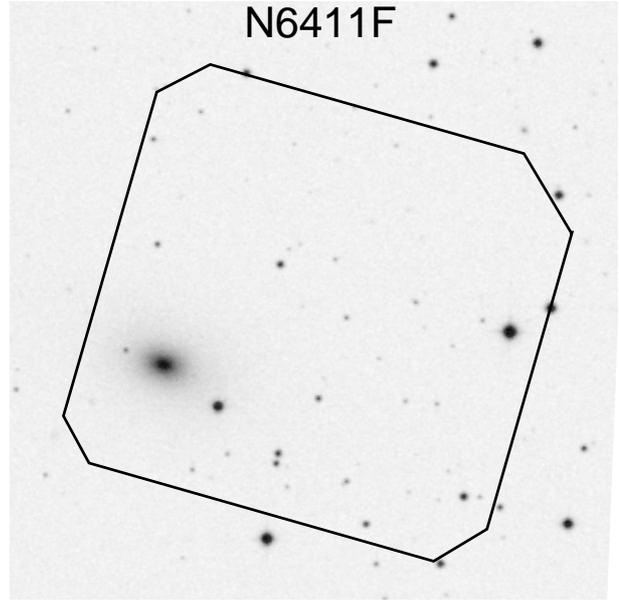}
 \caption{The GMOS field containing NGC\,6411, superimposed on an R image
from the Palomar Observatory Sky Survey. The image size is $8\times 
8\,{\rm arcmin^2}$. North is up, east to the left.}
 \label{dss}
\end{figure}

The data set consists of a field observed with the GMOS camera at Gemini 
North (programme GN-2015A-Q-70, PI: L.P. Bassino), containing the 
galaxy NGC\,6411 (hereafter N6411F, see Figure\,\ref{dss}), plus 
science verification observations from the Gemini Observatory Archive 
(GOA), which correspond to the programme GN-2001B-SV-67 (hereafter CompF) 
in order to estimate the field contamination. 
The GMOS field of view (FOV) is $\approx5.5\times5.5$\,arcmin.
Both programmes were observed in $g'$, $r'$ and $i'$ 
filters with similar exposure times (see Table\,\ref{obsdata}), using
$2\times2$ binning, which results in a scale of $0.146$\,arsec\,pixel$^{-1}$.
Considering the Galactic coordinates of the CompF ($l\approx80.5\degr$, 
$b\approx-52.7\degr$) and those of the N6411F ($l\approx89.7\degr$,
$b\approx 32.6\degr$), the foreground contamination is expected to be
small in comparison with that corresponding to background galaxies, and the
overall contamination is expected to be similar. In Section\,\ref{candsel} we 
detail this point.

 The reduction was carried out 
with the tasks from the GEMINI-GMOS package, within {\sc IRAF}. Around
20 bias exposures from nearby dates were used to built-up the 
master bias. Flat-fielding was performed using flats retrieved from
the GOA website. The reduction also included detector
mosaicking, image co-addition and the registration of the final images 
in the three filters. Basic data from the observations are listed in
Table\,\ref{obsdata}.

\begin{table}   
\begin{minipage}{85mm}   
\begin{center}   
\caption{Basic data from observations}    
\label{obsdata}   
\begin{tabular}{@{}lcccc@{}}   
\hline   
\multicolumn{1}{@{}c}{Name}&\multicolumn{1}{c}{Filter}&\multicolumn{1}{c}{Obs. date}&\multicolumn{1}{c}{Exp. time}&\multicolumn{1}{c}{$FWHM$}\\   
\multicolumn{1}{@{}c}{}&\multicolumn{1}{c}{}&\multicolumn{1}{c}{dd\,mm\,yyyy}&\multicolumn{1}{c}{s}&\multicolumn{1}{c}{$''$}\\   
\hline   
N6411F & $g'$ & $25\,05\,2015$ & $5\times780$ & $0.65$\\
 & $g'$ & $20\,06\,2015$ & $10\times780$ & $$\\
 & $r'$ & $25\,05\,2015$ & $9\times 450$ & $0.6$\\
 & $i'$ & $10\,06\,2015$ & $12\times300$ & $0.55$\\
CompF & $g'$ & $22\,08\,2001$ & $8\times900$ & $0.7$\\
 & $r'$ & $15\,11\,2001$ & $14\times600$ & $0.75$\\
 & $i'$ & $23\,08\,2001$ & $16\times300$ & $0.65$\\
\hline
\end{tabular}    
\end{center}    
\end{minipage}   
\end{table}   

\subsection{Photometry and point-source selection}
\label{photsec}
\begin{figure}    
\includegraphics[width=85mm]{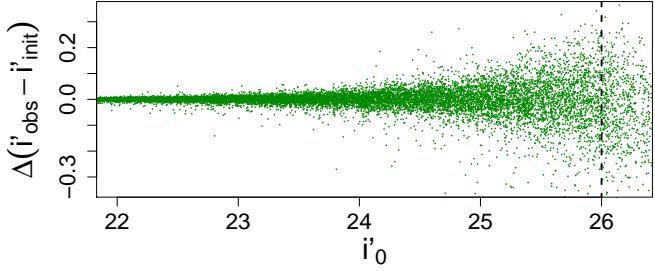}
\caption{Difference between the initial and final 
$i'_0$ magnitudes for artificial stars added to N6411F.
The dashed vertical line indicates completeness 
limit $i_0=26$\,mag chosen in Section\,\ref{complsec}.}    
\label{difi}    
\end{figure}

First, the galaxy light was subtracted from the GMOS 
original images of the N6411F in order to perform photometry on
a more or less flat background and detect sources that are 
immersed in the galaxy light. We applied to the images a 
square median filter of side 200\,px using FMEDIAN task from 
{\sc iraf}. This filtered image was subtracted from the original 
image, and the procedure was repeated a second time with a median
filter of side 40\,px to discard weak halos around objects near 
the galactic centre. Figure\,\ref{difi} shows
the difference between the initial and final $i'_0$ magnitudes for 
artificial stars added to the N6411F for completeness calculation.
The dashed vertical line indicates completeness limit $i_0=26$\,mag
chosen in Section\,\ref{complsec}. Similar comparisons were carried
out for the three filters in both fields, proving
that the procedure did not affect the point source photometry.

The software SE{\sc xtractor} \citep{ber96} was applied to the
$i'$ images to generate a source catalogue. This filter was
selected because it has the higher signal-to-noise ratio (S/N).
Considering that GCs usually have effective radii smaller 
than $R_{\rm eff}=10$\,pc \citep{bru12}, extragalactic GCs are
detected as point sources at the distance of NGC\,6411.
Hence, we used the stellarity index from SE{\sc xtractor} in 
order to identify extended sources. This parameter spans from
0 for resolved objects to 1 for ideal point sources. Sources with
lower values than $0.4$ were eliminated from the catalogue, 
following similar criteria from other extragalactic studies with
Gemini/GMOS \citep[e.g.][]{cas15a,esc15,bas17}.

\begin{figure}    
\includegraphics[width=85mm]{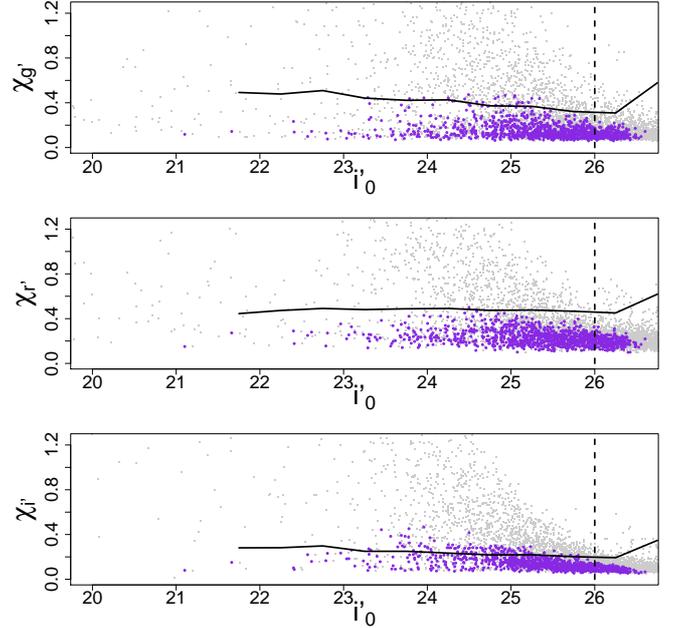}
\caption{$\chi^2$ parameter from PSF photometry as a function of $i'_0$ 
magnitudes for N6411F sources in the three filters. Objects fulfilling 
point source criteria are highlighted. The solid line represents in each
case the $\chi$ limit containing 90 per cent of the observed artificial
stars. In all panels the dashed vertical line indicates the completeness
limit $i_0=26$\,mag chosen in Section\,\ref{complsec}.}    
\label{chi}    
\end{figure}

The photometry was carried out with the DAOPHOT package \citep{ste87}
within {\sc iraf}. For each filter a second-order variable PSF was 
built-up from a sample of bright and relatively isolated stars, 
evenly distributed over the field. The final point-source catalogue 
was made considering the $\chi^2$ and sharpness parameters, from 
the task ALLSTAR. Figure\,\ref{chi} shows the $\chi^2$ parameter from 
the PSF photometry as a function of
$i'_0$ magnitudes for N6411F sources in the three filters. 
Objects fulfilling point 
sources criteria are highlighted. The solid line represents in each
case the $\chi^2$ limit containing $90\%$ of the observed artificial
stars. This analysis with both $\chi^2$ and sharpness parameters was
carried out for both fields, in order to corroborate the accuracy
of the selection criteria in discriminating point sources. The dashed 
vertical line indicates the completeness limit $i_0=26$\,mag chosen 
in Section\,\ref{complsec}.

\subsection{Photometric calibration}
For both programmes standard star fields from the $u'g'r'i'z'$ 
Standard Star Network \citep{smi02} have been observed during 
the same nights as the science observations. These fields contained 
between 6 and 12 bright stars measured in the three filters selected 
for our programme. Once we have determined from the curve-of-growth 
the proper aperture containing the total light, we fit the 
transformation equation: 
\begin{equation}
m_{std} = ZP + m_{inst} - K_{MK} (X-1) \nonumber
\end{equation}

\noindent with $m_{std}$ and $m_{inst}$ the standard and instrumental 
magnitudes, respectively, and $ZP$ are the photometric zero points. 
$K_{MK}$ is the mean atmospheric extinction at Mauna Kea, obtained from 
the Gemini Observatory Web 
Page\footnote{http://www.gemini.edu/sciops/instruments/gmos/calibration},
and $X$ the effective airmass. Then, the resulting zero points for the 
$g'$, $r'$
and $i'$ filters were $ZP_{g'}=3.06\pm0.03$, $ZP_{r'}=3.53\pm0.02$ and 
$ZP_{i'}=3.43\pm0.03$ in the case of the N6411F, and $ZP_{g'}=2.95\pm0.02$, 
$ZP_{r'}=3.17\pm0.03$ and $ZP_{i'}=2.91\pm0.03$ for the CompF.

Aperture corrections were calculated from the stars selected for each PSF, 
and extinction corrections were obtained from \citet[][available in NED]{sch11}
values.

\subsection{Photometric completeness}
\label{complsec}
\begin{figure}
 \includegraphics[width=80mm]{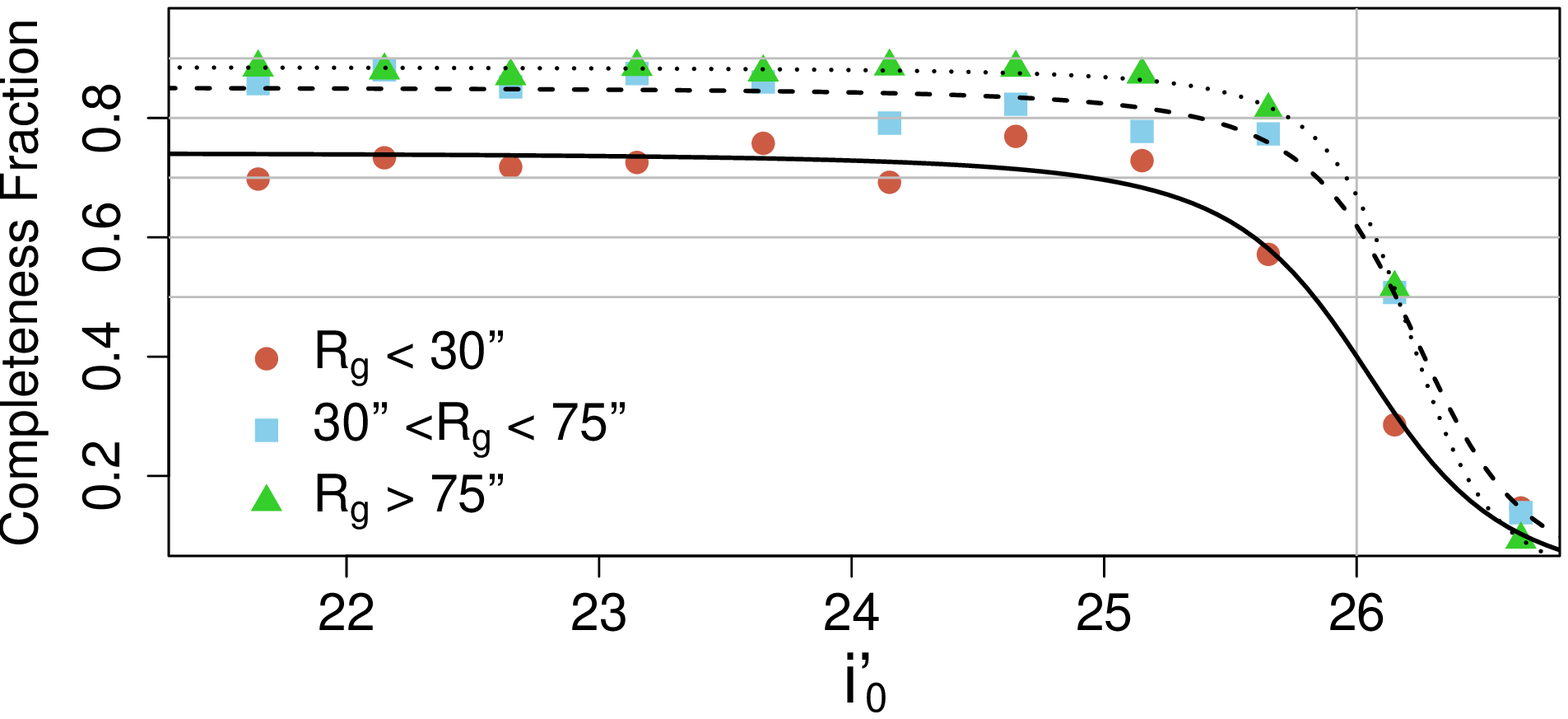}\\
 \includegraphics[width=80mm]{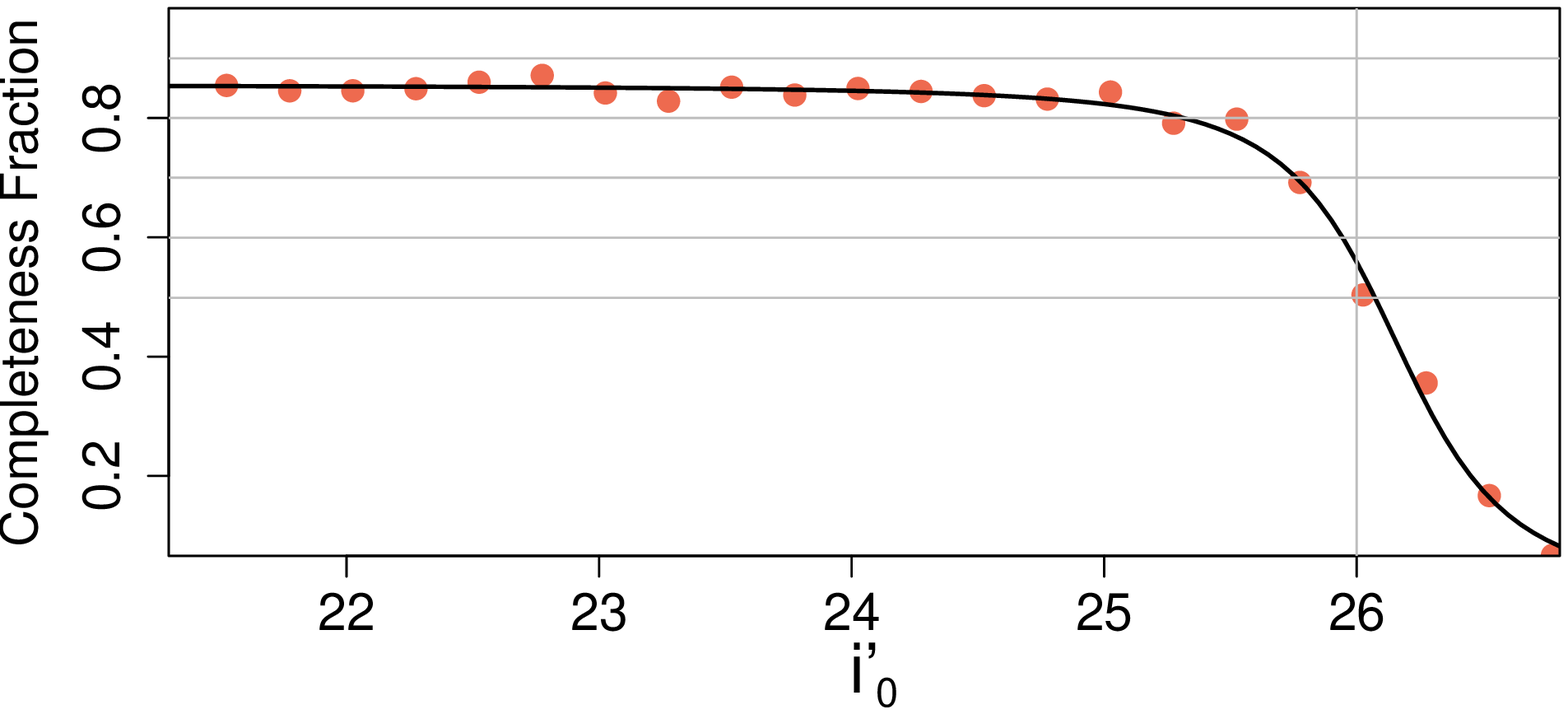}\\
\caption{Completeness curves for N6411F (upper panel) and CompF 
(lower panel). In the first case, the artificial stars were split
in three radial ranges to quantify the effect of the noise increase
towards the galaxy centre.} 
 \label{comp}
\end{figure}

In order to estimate the photometric completeness for both fields, 
we added 250 artificial stars in the three filters, evenly distributed 
over the entire field, and with colours in the expected ranges for GCs. 
Their photometry was carried out in the same way as for the original fields,
and the process was repeated 80 times, achieving a sample of 20\,000
artificial stars. 
The upper panel of Figure\,\ref{comp} shows the completeness for the 
N6411F, split in three galactocentric regimes to take into account the 
decline of the completeness towards the galaxy centre. We chose as the 
magnitude limit $i'_0=26$\,mag, where the completeness falls below $40\%$
for point sources at  $10'' < R_g < 30''$ and $60\%$ for the outer radii.
At the magnitude limit, the completeness for the CompF falls below 
$50\%$. In order to apply completeness corrections in our analysis,
we fitted to each curve an analytic function of the form:

\begin{equation}
f(m) = \beta \left( 1 - \frac{\alpha(m-m_0)}{\sqrt{1+\alpha^2(m-m_0)^2}}\right) 
\end{equation}

\noindent similar to that used by \citet{har09c}, with $\beta$, $\alpha$
and $m_0$ free parameters.

The comparison between the initial brightness of artificial stars and 
their subsequent photometry (shown in Figure\,\ref{difi} for a single
filter) proved that the application of the median filter does not modify 
the point source photometry.

\section{Results}

\subsection{NGC\,6411 surface photometry}
\label{galsec}

We obtained the NGC\,6411 surface brightness profiles in $g'_0$ and
$i'_0$, and the colour profile in $(g'-i')_0$ applying the
task ELLIPSE, within {\sc iraf}. 

\begin{figure}    
\includegraphics[width=85mm]{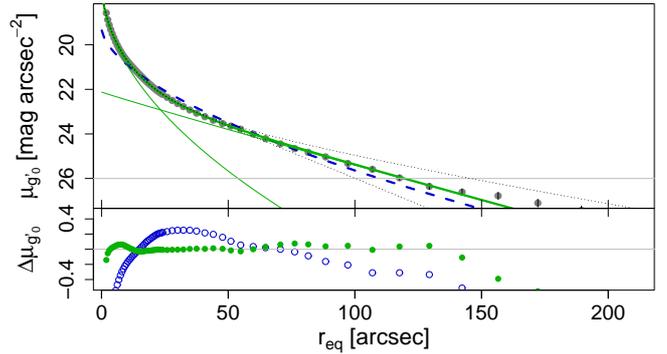}
\caption{NGC\,6411 surface brightness profile in the $g'$ filter. 
The blue dashed line represents a single S\'ersic model fitted
to the galaxy. Its residuals are plotted in the lower panel with 
blue open symbols. The green and thin solid lines correspond to
two S\'ersic models, while the thick one indicates the sum of both
components. The composite residuals are plotted in the lower panel
with green filled symbols.}   
\label{prof}    
\end{figure}

The upper panel of Figure\,\ref{prof} shows the surface brightness 
profile of the galaxy in the $g'$ filter. The  blue dashed line
represents a single S\'ersic model \citep{ser68} fitted to the galaxy.
The result presents large residuals, plotted in the lower panel with
blue open symbols. The green and thin solid lines correspond
to two S\'ersic models, while the thick one indicates their sum. The
composite residuals are plotted in the lower panel with green
filled symbols, resulting in a more accurate fit.
The dotted curves represent the change in the surface brightness profile 
of the galaxy when the sky level varies an amount equal to its
Poisson noise.

The adopted form for the S\'ersic profile was:

\begin{equation}  
\mu(r) = \mu_0 + 1.0857\,\left(\frac{r}{r_0}\right)^{\frac{1}{n}},  
\end{equation}  
  
\noindent where $\mu_0$ is the central surface brightness, $r_0$ 
is a scale parameter and $n$ is the S\'ersic shape index. The
parameters are listed in Table\,\ref{sersic}, together with the
effective radii, derived from the relation

\begin{equation}  
r_{eff} = b_{\rm n}^{n}r_0 
\end{equation}  

\noindent where $b_n$ is a function of $n$ index, which might be
estimated with the expression given by \citet{cio91}.
From the surface brightness profiles in both filters we obtained 
${\rm R_{eff}}\approx 6.4$\,kpc. This value is twice the half-light radius
derived by \citet{gon15} in the CALIFA survey.

\begin{table}   
\begin{minipage}{85mm}   
\begin{center}   
\caption{Parameters of the S\'ersic models fitted to the $g'$ surface
brightness profile, with $r_0$ and $r_{\rm eff}$ expressed in arcsec.}    
\label{sersic}   
\begin{tabular}{@{}lcccc@{}}   
\hline   
\multicolumn{1}{@{}c}{Component}&\multicolumn{1}{c}{$\mu_{0}$}&\multicolumn{1}{c}{$r_0$}&\multicolumn{1}{c}{$n$}&\multicolumn{1}{c}{$r_{\rm eff}$}\\   
\hline   
Single & $19.25\pm0.29$ & $5\pm1.6$ & $1.69\pm0.15$ & $32.9$\\
Inner & $17.14\pm0.08$ & $0.88\pm0.08$ & $1.96\pm0.04$ & $10.8$\\
Outer & $22.1\pm0.40$ & $31.8\pm4.5$ & $1.04\pm0.10$ & $56.7$ \\
\hline
\end{tabular}    
\end{center}    
\end{minipage}   
\end{table}   

\begin{figure}    
\includegraphics[width=85mm]{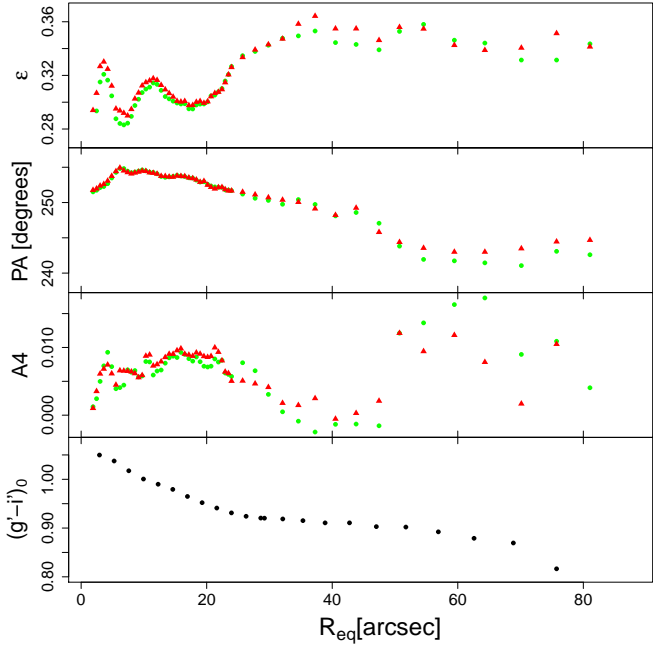}\\    
\caption{From top to bottom, ellipticity, position angle, harmonic A4
and $(g'-i')_0$ colour profile, obtained from our observations. In the 
three upper panels, green circles and red triangles represent parameters 
from $g'$ and $i'$ filters, respectively.}    
\label{param}    
\end{figure}

Figure\,\ref{param} shows the 
radial dependence of ellipticity ($\epsilon$), position angle and the 
higher harmonic A4, derived for $g'$ (green circles) and $i'$ (red triangles)
filters, and the $(g'-i')_0$ colour profile. The four panels are plotted
in terms of the equivalent radius, with $r_{eq}= a \sqrt{1-\epsilon}$, where
$a$ is the isophote semimajor axis. No striking difference is 
seen among them in any of these morphological parameters. 
The parameter A4 is mainly positive, which is expected in disky isophotes,
but deviation from zero hardly exceed $0.01$ in the inner 50\,arcsec. 
The $(g'-i')_0$ colour profile was obtained from the galaxy colour map.
It presents a gradient, becoming bluer towards the outskirts. The 
integrated colour resulted in $(g'-i')_0=0.92$. The CALIFA survey 
\citet{gon15} used two different single stellar populations (SSPs) 
datasets to analyse their data. In both cases they found
NGC\,6411 to fit a SSP with an intermediate-age population of $3-4$\,Gyr 
and suprasolar metallicity, $\log_{10}(Z/Z_{\odot})=0.14-0.25$. 
From the webtool CMD\,3.0\footnote{
http://stev.oapd.inaf.it/cmd}, assuming \citet{bre12} isochrones and 
Chabrier IMF we obtained $(g'-i')=1.05-1.15$. This is redder 
than our colour profile, 
but the CALIFA field is smaller than the effective radius.

There is a change in the behaviour of $\epsilon$, A4 and 
the colour profile around $\approx 25$\,arcsec. 
This corresponds to the galactocentric distance where the 
disky outer component starts to 
dominate over the inner spheroidal one in Figure\,\ref{prof},
lending additional support to our choice of the two-component fit.

\subsection{GC candidates selection from colours}
\label{candsel}
\begin{figure}    
\includegraphics[width=80mm]{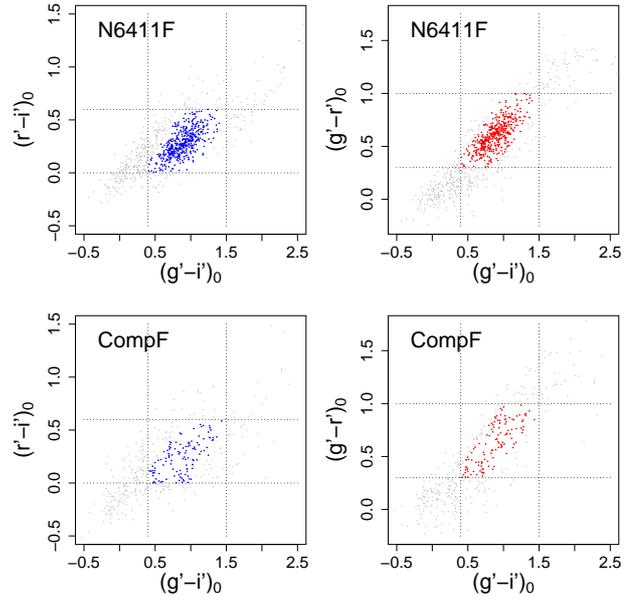}    
\caption{Colour-colour diagrams for both fields. Dotted lines indicate
the colour ranges of GC candidates (see Section\,\ref{candsel}).}   
\label{dcc}    
\end{figure}

Figure\,\ref{dcc} shows $(r'-i')_0$ vs. $(g'-i')_0$ and $(g'-r')_0$
vs. $(g'-i')_0$ colour-colour diagrams (CCD) for point sources in
both fields. Because old GCs present narrow colour ranges, CCD can 
be helpful to distinguish them from contamination \citep[e.g.][]{fai11}.
Hence, the GC candidates where selected from the point sources ranging
$0.4 < (g'-i')_0 < 1.4$, $0.3 < (g'-r')_0 < 1$ and $0 < (r'-i')_0 < 0.5$.
These values are similar to those used in previous GCS studies 
\citep[e.g.][]{cas15a,esc15,bas17}. There are 550 GC candidates brighter 
than our completeness limit in N6411F and 170 point sources in CompF
complying with the same colour constraints. We used the Besan\c{c}on
models \citep{rob03} to simulate the contribution of foreground stars
with colours in the ranges previously indicated and magnitudes between 
$i'=22$ and $i'=26$\,mag in both fields. For an area of 1\,$deg^2$
centred on the Galactic coordinates of CompF we obtained 2238 stars,
implying an expected number of $\approx 19$ stars for the GMOS FOV. It 
was also simulated an area centred on the N6411F coordinates, resulting
in 3623 stars and $\approx 30$ considering the GMOS FOV. Then, the
contamination is dominated by background galaxies instead of foreground
stars.

\begin{figure}    
\includegraphics[width=80mm]{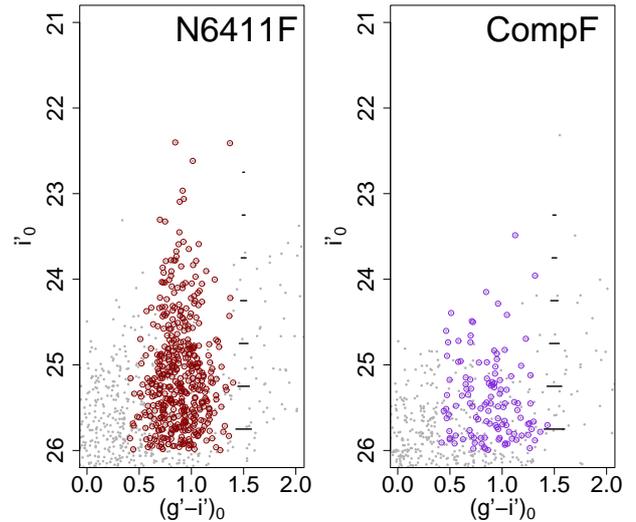}    
\caption{Colour magnitude diagrams for both fields. Point sources that
fulfill the colour criteria are highlighted. Horizontal bars indicate the mean 
$(g'-i')_0$ uncertainties}   
\label{dcm}    
\end{figure}

The colour magnitude diagrams (CMD) for point sources are shown in 
Figure\,\ref{dcm}, with those sources fulfilling the colour criteria
highlighted. Horizontal bars indicate the mean $(g'-i')_0$ uncertainties for
GC candidates. Comparing both fields, the objects distribution outside the 
GCs colour range is similar, which give confidence to the point sources 
selection described in Section\,\ref{photsec}.

\newcommand*{\MyIndent}{\hspace*{0.6cm}}
\begin{table*}   
\begin{minipage}{150mm}   
\begin{center}   
\caption{Parameters of the Gaussians fitted to the colour distribution 
for different radial ranges. The columns indicate the colour peak
and Gaussian width for blue and red GCs, respectively, the fraction of
red GCs ($F_{red}$) and the $DD$ parameter from {\sc gmm} analysis. We 
also indicated the kurtosis ($\kappa$) in each case.}    
\label{fit}   
\begin{tabular}{@{}lcccccc@{}}   
\hline   
\multicolumn{1}{@{}c}{}&\multicolumn{1}{c}{$(g'-i')_{0,blue}$}&\multicolumn{1}{c}{$\sigma_{blue}$}&
\multicolumn{1}{c}{$(g'-i')_{0,red}$}&\multicolumn{1}{c}{$\sigma_{red}$}&\multicolumn{1}{c@{}}{$F_{red}$}&\multicolumn{1}{c@{}}{$DD$}\\   
\hline   
\multicolumn{1}{c}{}&\multicolumn{5}{c}{Entire Sample | $\kappa=-0.89\pm0.04$}&\multicolumn{1}{c}{}\\   
\hline   
Unimodal & $0.90\pm0.01$ & $0.14\pm0.01$ & $-$ & $-$ & $-$ & $-$\\
Bimodal & $0.81\pm0.02$ & $0.09\pm0.01$ & $1.04\pm0.02$ & $0.07\pm0.01$ & $0.38\pm0.07$ & $2.92\pm0.30$\\
\hline
\multicolumn{1}{c}{}&\multicolumn{5}{c}{$10''< R_{gal} < 40''$ | $\kappa=-1.02\pm0.01$}&\multicolumn{1}{c}{}\\
\hline
Unimodal & $0.91\pm0.02$ & $0.15\pm0.01$ & $-$ & $-$ & $-$ & $-$\\
Bimodal & $0.80\pm0.02$ & $0.09\pm0.01$ & $1.04\pm0.03$ & $0.08\pm0.01$ & $0.45\pm0.14$ & $2.93\pm0.41$\\
\hline    
\multicolumn{1}{c}{}&\multicolumn{5}{c}{$40''< R_{gal} < 80''$ | $\kappa=-0.89\pm0.02$}&\multicolumn{1}{c}{}\\
\hline    
Unimodal & $0.89\pm0.02$ & $0.13\pm0.01$ & $-$ & $-$ & $-$ & $-$\\
Bimodal & $0.83\pm0.03$ & $0.10\pm0.01$ & $1.04\pm0.03$ & $0.05\pm0.01$ & $0.31\pm0.11$ & $2.63\pm0.43$\\
\hline   
\multicolumn{1}{c}{}&\multicolumn{5}{c}{\MyIndent $R_{gal} > 80''$ | $\kappa=-0.93\pm0.13$}&\multicolumn{1}{c}{}\\
\hline   
Unimodal & $0.88\pm0.02$ & $0.13\pm0.01$ & $-$ & $-$ & $-$ & $-$\\
Bimodal & $0.78\pm0.02$ & $0.08\pm0.01$ & $1.02\pm0.02$ & $0.07\pm0.02$ & $0.35\pm0.13$ & $3.34\pm0.65$\\
\hline    
\end{tabular}    
\end{center}    
\end{minipage}   
\end{table*}   

\subsection{Colour distributions}
\label{colsec}

\begin{figure}    
\includegraphics[width=85mm]{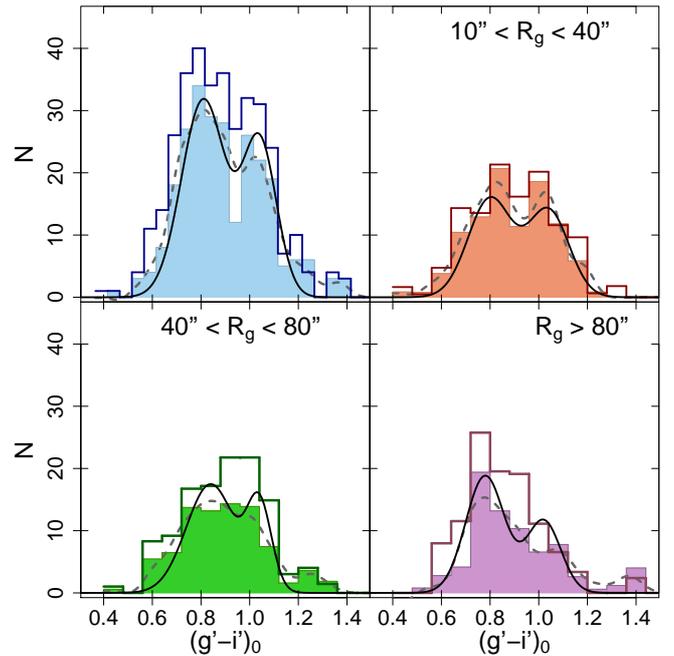}    
\caption{Background corrected colour distribution for all the GC candidates 
(upper left panel), and split in three radial regimes. The filled histograms
correspond to GC candidates brighter than $i'_0= 25.5$\,mag, the limit 
applied for {\sc gmm} analysis, while the open ones take into account GC
candidates up to $i'_0= 26$\,mag, the completeness limit. The dashed curves 
indicate smoothed distributions obtained with a Gaussian kernel. 
The solid curves represent the {\sc gmm} results.}    
\label{dcol}    
\end{figure}

The filled histograms in Figure\,\ref{dcol} show the background 
corrected colour distribution for all GC candidates brighter than $i'_0=
25.5$\,mag (upper left panel), and split in three radial regimes (other 
panels). The bin width is $\Delta(g'-i')= 0.05$ for the full sample
histogram and $\Delta(g'-i')= 0.08$ for each of the three subsample 
histograms. The dashed curves 
indicate smoothed distributions obtained with a Gaussian kernel for all
GCs in each radial regime.

From visual inspection, bimodality is not so evident in the split
samples. In order to determine whether the GCS colour distribution is 
better represented by two Gaussians instead of one, we used the Gaussian 
Mixture Method algorithm \citep[{\sc gmm},][]{mur10}. 
For each sample we eliminated the GC candidates with colours close to 
those of point-sources in the CompF, up to the expected number 
of contaminants. We repeated this procedure 25 times, in order to obtain 
a set of samples free of contamination and we applied {\sc gmm} to each one 
of them. When we considered all GC candidates brighter than $i'_0=26$\,mag, 
the {\sc gmm} results differed significantly, presenting a large dispersion 
in the colour peaks. In a second iteration we chose $i'_0=25.5$\,mag as the 
magnitude limit, implying colour uncertainties below 0.05\,mag. This latter 
selection shortened the samples size, but {\sc gmm} results 
remained stable in the different runs.

In order to investigate whether the colour uncertainties triggered the 
{\sc gmm} results, we simulated samples with 120 $(g'-i')_0$ colours 
drawn from two Gaussians distributions, centred at  $(g'-i')_0=0.9$\,mag and 
$(g'-i')_0=1.1$\,mag, with $\sigma= 0.12$\,mag. We added to these error-free 
colours a noise component represented by a third
Gaussian with mean 0.075\,mag and dispersion $0.015$\,mag. This latter
function was obtained from the $(g'-i')_0$ uncertainties analysis for GC
candidates with $25.5 < i'_0 < 26$. The results from applying {\sc gmm} to 100
artificial samples presented a large scatter in the parameters. The peaks of
the two Gaussians span  $0.7-0.99$\,mag and $1.04-1.30$\,mag, respectively. 
The Gaussians dispersions span $0.04-0.15$\,mag. and the fraction of red GCs
varies from $15\%$ to $80\%$. This justifies the exclusion of  GC candidates 
fainter than $i'_0=25.5$\,mag from the {\sc gmm} analysis.

This is also evident when we compare the colour distribution obtained
when the faintest limit is chosen at $i'_0=26$\,mag, corresponding to the
open histograms in Figure\,\ref{dcol}. In all cases the bimodality seems to
be blurred by colour uncertainties.

The mean values obtained from {\sc gmm} for the selection of GC candidates 
brighter than $i'_0=25.5$\,mag are listed in Table\,\ref{fit}.

In all cases the kurtosis ($\kappa$) is negative, pointing to a 
flattened (platykurtic) distribution. This is a necessary but not 
sufficient condition of bimodality.
The $DD$ parameter (related to the distances between the means of 
the Gaussians in terms of their dispersions) is above 2, the minimum 
value expected when the distribution is better described by the sum of two 
Gaussians. This is fulfilled even considering the uncertainties in the $DD$
calculus. The fraction of red GCs ($F_{red}$) is $\approx 40\%$, similar 
to other giant ellipticals in denser environments \citep[][and references 
therein]{cas17,har15}. When $F_{red}$ is obtained from the integration of the 
Hubble profiles for both subpopulations, 
the result is $\approx 40\%$, in agreement with the fraction
derived from the colour distribution. Assuming its $V$ absolute magnitude is 
$M_V\approx -21.5$ (see Section\,\ref{lfsec}), the colour peaks are in 
agreement with those expected from literature compilations of early-type 
galaxies in the same photometric system \citep[e.g.][]{fai11}.
 
From the evolution of $F_{red}$ along the radial regime, red GCs seem to be 
slightly more concentrated towards the galaxy centre, with lower values of
$F_{red}$ in the outer bins. The colour peaks and dispersions do not vary 
significantly with galactocentric distance.

\begin{figure*}    
\includegraphics[width=170mm]{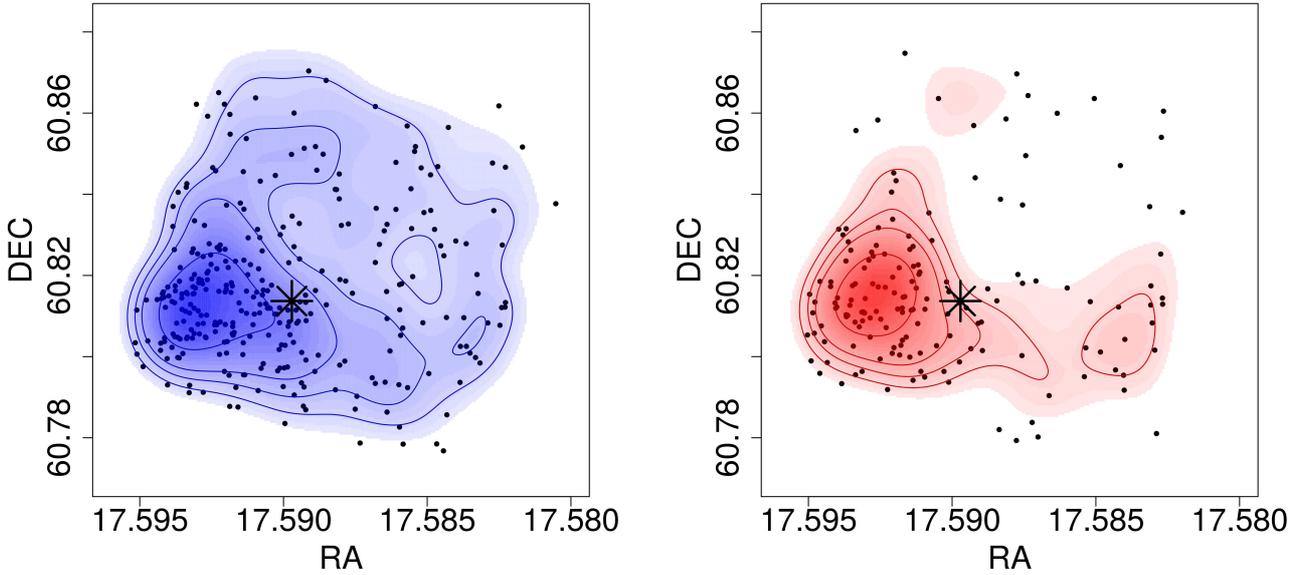}    
\caption{Projected spatial distribution for blue (left panel) and 
red (right panel) GC candidates, asuming $(g'-i')_0=0.95$ as the 
colour limit. The field of view is $5.5\times5.5\,{\rm arcmin}^2$. 
North is up, and East to the left. The asterisk indicates
the projected position of the SNI\,1999da. The colour palette
correspond to the smoothed distribution, applying a Gaussian kernel.
The contour curves represent density levels corresponding to 30, 40,
50 and $75\%$ of the maximum.}
\label{espa}    
\end{figure*}

\subsection{Spatial and radial distributions}
\label{radsec}
The projected spatial distribution for GC candidates is shown in
Figure\,\ref{espa}. The left panel corresponds to blue GCs and the 
right panel to red ones, assuming $(g'-i')_0=0.95$ as the colour 
limit between them. Blue GC candidates seem to dominate. The asterisk
indicates the projected position of the SNI\,1999da.

The spatial distribution might deviate from azimuthal symmetry, but it 
is not possible to perform an analysis of the GCS azimuthal distribution 
due the presence of a saturated star $\approx 0.9$\,arcmin to the West 
of NGC\,6411. In order to go further with the implications of this
uncertainty we fitted the projected radial distribution of the entire
population of GCs using concentric circular rings and elliptical rings.
These latter ones were defined assuming the PA and $\epsilon$ derived
for the galaxy in Section\,\ref{galsec}.

\begin{figure}    
\includegraphics[width=85mm]{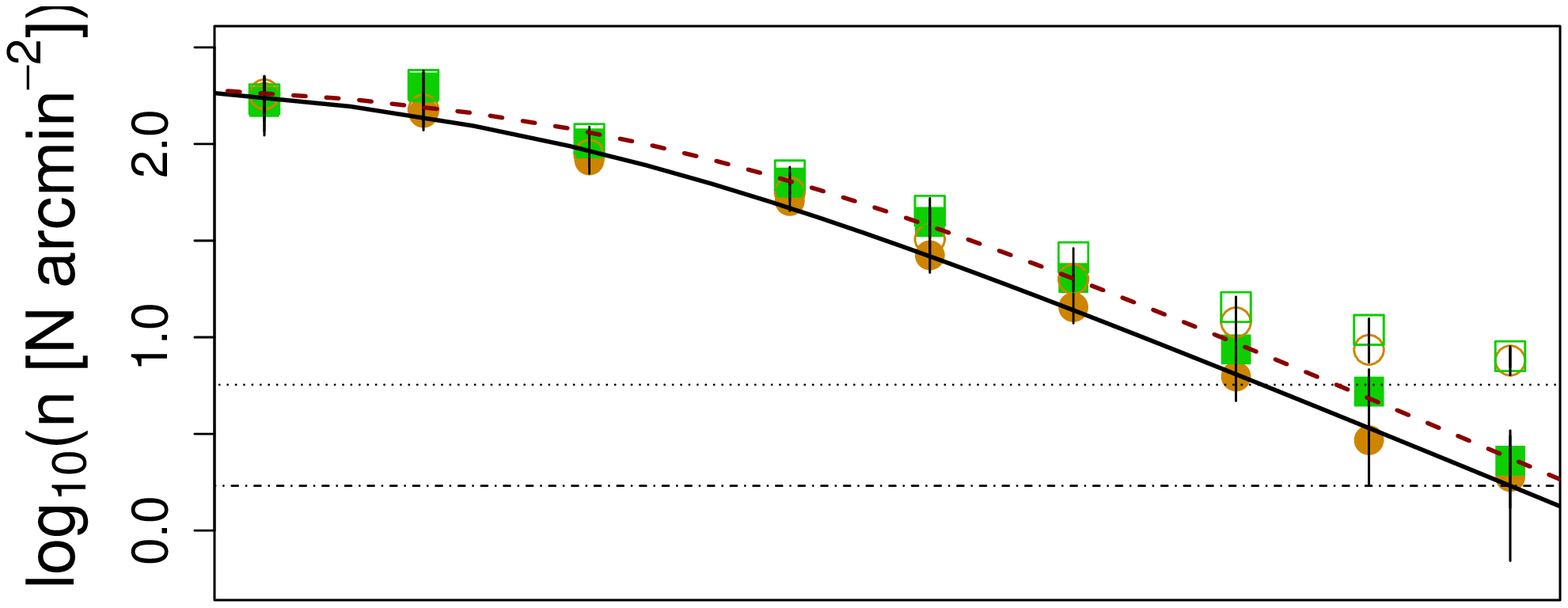}    
\includegraphics[width=85mm]{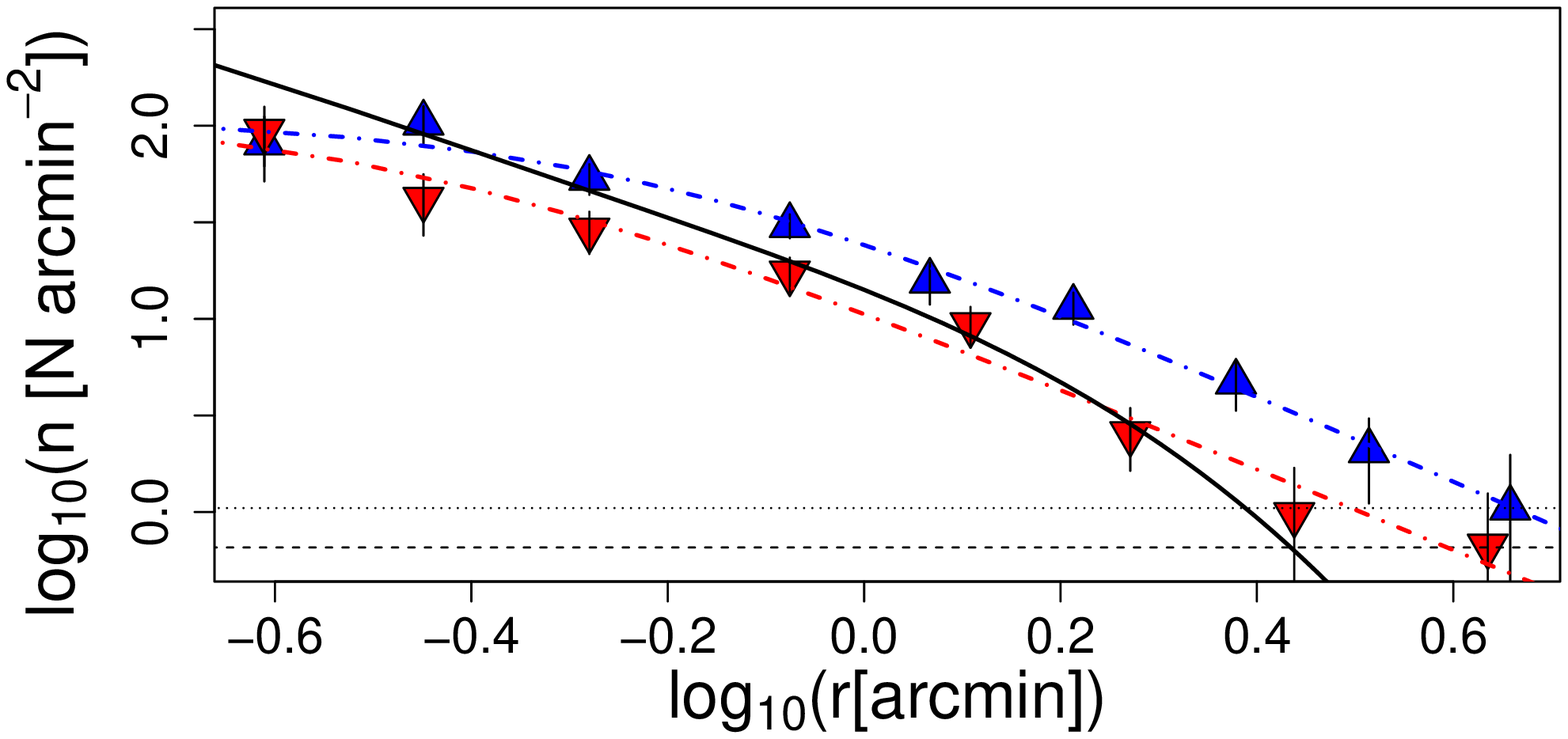}    
\caption{{\bf Upper panel:} radial projected distribution for GC candidates, 
background and 
completeness corrected, obtained from concentric circles (orange circles)
or ellipses with the same ellipticity and position angle than the
galaxy profile (green squares). The solid curve corresponds to the Hubble 
modified profile fitted to the radial distribution with circular rings, 
and dashed one corresponds to the elliptical ones. The horizontal dotted 
and dashed lines show the background level and its $30\%$, which is used 
to define the GCS extension. 
{\bf Lower panel:} radial projected distribution
for GC candidates split in blue (upwards triangles) and red (downwards
triangles) subpopulations (see Section\,\ref{colsec}). Solid lines represent
the Hubble modified fits, while dotted and dashed lines represent the 
value used to define the extension of the populations for blue and red GCs,
respectively. The black solid curve represents the galaxy luminosity profile
in $g'$ filter, arbitrary scaled.}    
\label{drad}    
\end{figure}

The upper panel of Figure\,\ref{drad} shows the radial distribution for 
GC candidates brighter than $i'_0 = 26$. Orange open (filled) circles 
correspond to the radial profile with circular rings once it was corrected 
by completeness (completeness and contamination), and green squares 
are analogues when elliptical rings are used. The horizontal dotted 
and dashed-dotted lines represent the background level and the $30\%$ of 
its value, respectively. This latter value has been used in past studies 
to define the GCS extension \citep[e.g.][]{cas15a,bas17}. 
 In order to better describe the distribution flattening towards the 
centre of the galaxy, we fitted a modified Hubble profile 
\citep[e.g.][]{bin87,dir03b} of the form:

\begin{equation}
n(r) = a\left(1 + \left(\frac{r}{r_0}\right)^2\right)^c
\end{equation}

\noindent which behaves as a power-law with index $2c$ when $r >> r_0$.
The parameters obtained for circular and elliptical rings are indicated in
Table\,\ref{trad} and correspond to solid and dashed curves, respectively.

\begin{table}   
\begin{minipage}{85mm}   
\begin{center}   
\caption{Parameters of the Hubble modified law fitted to the GC 
candidates radial profile, in the cases of circular and elliptical 
rings.}    
\label{trad}   
\begin{tabular}{@{}lccc@{}}   
\hline   
\multicolumn{1}{@{}c}{Case}&\multicolumn{1}{c}{$a$}&\multicolumn{1}{c}{$r_0$}&\multicolumn{1}{c}{$c$}\\   
\hline
\multicolumn{4}{c}{Elliptical symmetry}\\
\hline
Entire pop. & $221\pm26$ & $0.58\pm0.09$ & $-1.09\pm0.08$\\
\hline
\multicolumn{4}{c}{Circular symmetry}\\
\hline
Entire pop. & $233\pm28$ & $0.44\pm0.06$ & $-1.05\pm0.05$\\
Blue subpop.& $112\pm17$ & $0.59\pm0.11$ & $-1.13\pm0.10$\\
Red subpop.&  $123\pm30$ & $0.33\pm0.15$ & $-1.05\pm0.15$\\
\hline
\end{tabular}    
\end{center}    
\end{minipage}   
\end{table}   

The extension of the GCS is $\approx 5$\,arcmin, i.e. $\approx 65$\,kpc
at the distance of NGC\,6411 (see Section\,\ref{lfsec}) for the circular 
case, and
marginally larger, $\approx 70$\,kpc, in the elliptical one. There were no 
significant differences in the fitted parameters, nor in the extension of the
system for both geometrical schemes. Hence, for simplicity in the following 
we refer to the circular case for the radial profile.

We also fitted the radial projected distributions split in
subpopulations (lower panel in Figure\,\ref{drad}). We assumed $(g'-i')_0=0.95$
as the limiting colour, which is in agreement with colour distributions from
Section\,\ref{colsec}. The distribution for blue GCs is represented with
upwards triangles and downwards triangles were used for the red ones. 
Solid lines represent the fits of the modified Hubble profile,
which are also listed in Table
\ref{trad}. The $c$ parameter is slightly larger for red GCs, but 
differences are within the uncertainties. Dotted and dashed lines represent 
the $30\%$ of the background for blue and red GCs, respectively, used to
determine the extension for each subpopulation. In both cases it 
reaches similar galactocentric
distances, but blue GCs present larger densities at all radii, which explains
their more disperse distribution in Figure\,\ref{espa}.
 Hence, the radial distribution for red GCs deviates from the usual 
finding in bright ellipticals, where they seem to be more concentrated towards 
the galaxy centre. 
The black solid curve represents the galaxy luminosity profile
in $g'$ filter, arbitrarily  scaled for comparison purposes. Both GC 
subpopulations are more extended than the galaxy profile.

\subsection{Luminosity function and GCs population}
\label{lfsec}

\begin{figure}    
\includegraphics[width=85mm]{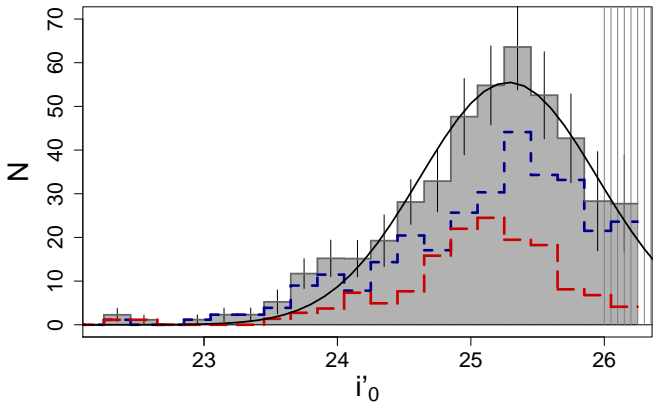}\\    
\caption{Background and completeness corrected luminosity 
function (GCLF) for GCs. The errorbars assume Poisson 
uncertainties for science and background measurements, and
the binwidth is 0.2.The short and long dashed histograms
correspond to blue and red GC subpopulations, respectively.
The vertical lines indicate the 
luminosity range avoided due to completeness drop.}    
\label{lf}    
\end{figure}

Figure\,\ref{lf} shows the GC luminosity function (GCLF) 
background and completeness corrected, over the entire field. 
We assumed Poisson uncertainties for the errorbars in both, 
science and background measurements. The binwidth is 0.2\,mag. 
The short and long dashed histograms indicate the GCLFs for blue 
and red GCs, respectively, assuming $(g'-i')=0.95$ as the colour
limit. The grey vertical lines 
indicate the range
excluded because of  the declining completeness.
The solid curve represents the Gaussian profile fitted to the
GCLF, which resulted in a turn-over magnitude (TOM) and width 
of $i'_{0,TOM}=25.29\pm0.05\pm0.1$ and $\sigma=0.68\pm0.05\pm0.1$, 
where the second uncertainty comes from the binwidth.

Old GC populations usually present a Gaussian GCLF, with a 
nearly universal TOM in the $V$-band of $M_{V,TOM}\approx-7.4$
\citep[e.g.][]{ric03,jor07}. From the unimodal distributions
in Table\,\ref{fit}, GC candidates brighter than $i_0'=26$ 
present a mean colour $(g'-i')\approx0.9$. If we assume
Equations\,1 and 2 from \citet{bas17} in order to transform
our $i'_{0,TOM}$ magnitude, the TOM in the $V$ 
filter is $V_{0,TOM}= 25.9\pm0.15$. Hence, the distance 
modulus results $(m-M)\approx33.3\pm0.15$, in agreement with 
measurements from SBF and the fundamental plane from  
\citet{bla01}, $(m-M)\approx32.9\pm0.35$ and $(m-M)\approx33.4\pm0.4$,
respectively, but slightly smaller than SNIa measurements \citep{tul13},
$(m-M)\approx33.6\pm0.1$.
Despite the larger noise when we subdivide the GCLF into 
subpopulations, the one corresponding to red GCs (dotted
histogram in Fig.\,\ref{lf}) might present a brighter TOM and 
a lower dispersion.

For magnitudes brighter than $i'_0=24$\,mag, there seems to be an 
excess of GC candidates in comparison with fainter ones. This
issue will be addressed in the following Section.

From the set of parameters fitted to the GCLF we found that 
the fraction of GCs brighter than $i'_0=26$\,mag is 
$0.85\pm0.02$. Due to the lack of areal coverage, we decided to 
integrate the radial profile instead of the GCLF to obtain the total
population of GCs. Then, from the numerical integration of 
the modified Hubble profiles derived in Section\,\ref{radsec} we obtained a
population of $610\pm22$ GCs brighter than $i'_0=26$ up to a 
galactocentric distance of 5\,arcmin for circular symmetry, and 
$580\pm26$ GCs up to $5.5$\,arcmin in the elliptical case. 
Then, the total population of GCs results $720\pm40$ and $685\pm45$, 
respectively. Adopting the total $V$ magnitude from \citet{dev91}, 
$V_{tot}= 11.85\pm0.13$, the foreground 
extinction from \citet{sch11} and the distance modulus previously 
derived, $m-M= 33.3\pm0.15$, the absolute magnitude of the galaxy results 
$M_V=-21.6\pm0.24$. Then, the specific frequency\footnote{The specific 
frequency $S_N$ or number of clusters per unit galaxy luminosity is 
defined as $S_N=N_{GCs}\times10^{0.4(M_V+15)}$} \citep{har81}
$S_N\approx1.65\pm 0.45$, where the uncertainty is dominated by the 
propagation of the distance error.

\subsection{An excess of bright GCs?}
\label{bright}

With a strict photometric $\chi^2$-selection, we may miss slightly extended 
objects, whose effective radii are between bright GCs and ultra compact dwarfs
\citep[e.g.][]{nor14,bas17}. Hence, we decided to relax the $\chi^2$ limits in
the three filters in order to analyse the presence of bright GCs. This
significantly enlarges the sample of GC candidates brighter than 
$i'_0= 24$\,mag, but results in a marginal variation for fainter GC candidates 
($\approx 7\%$). Despite of this, in the following a 
comparison between faint and bright GC candidates will be performed for this
sample obtained with relaxed $\chi^2$ limits.
Some objects with $\chi^2$ larger than typical values for artificial stars 
(see Figure\,\ref{chi}) deviate from point sources although the visual 
inspection of their radial profiles does not show evidence of saturation and 
their counts are not in the CCD non-linearity range.

\begin{figure}    
\includegraphics[width=85mm]{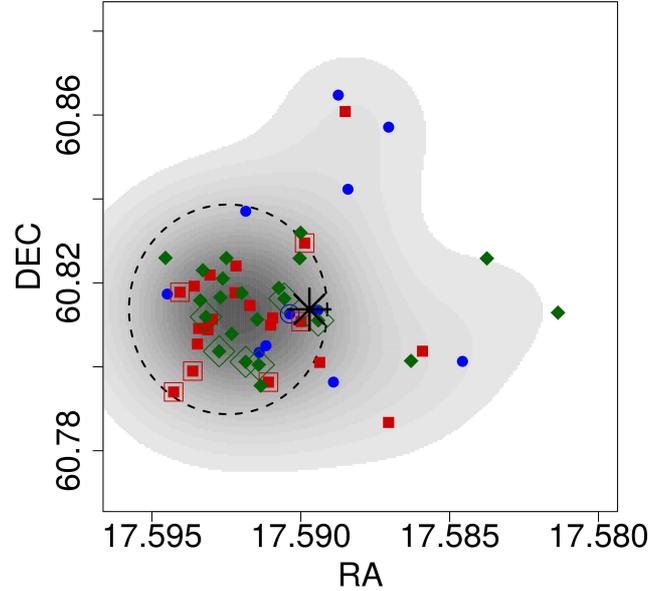}\\    
\caption{Projected spatial distribution for candidates brighter than 
$i'_0= 24$\,mag and with typical colours  of GCs. North is up, East to 
the left. The dashed circle is centred in the galaxy and its radius is
90\,arcsec. The grey scale indicates the smoothed projected density, and 
the symbols discriminate objects bluer than $(g'-i')_0=0.9$ (blue
circles), redder than $(g'-i')_0=1$ (red squares), and with intermediate
colours (green diamonds).The grey scale indicates the smoothed projected 
density.}    
\label{espa3}    
\end{figure}

\begin{figure}    
\includegraphics[width=85mm]{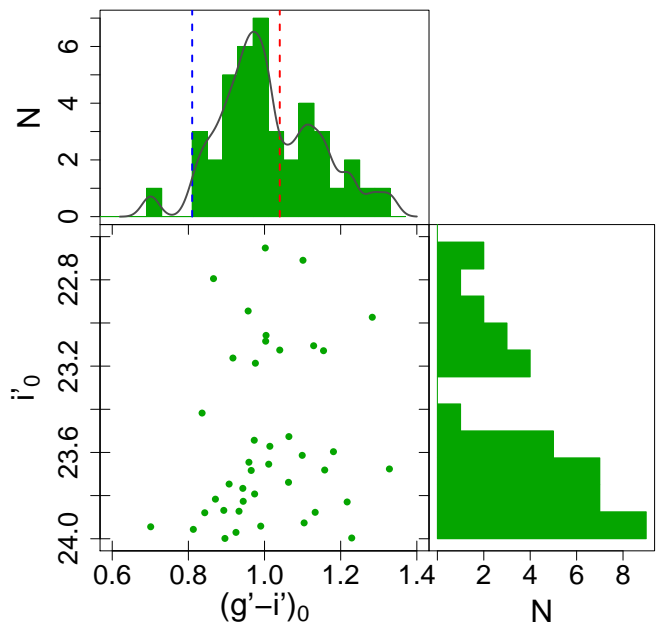}\\    
\caption{{\bf Bottom left panel:} CMD for those objects brighter than 
$i'_0= 24$\,mag with galactocentric distances up to 90\,arcsec.
{\bf Bottom right panel:} luminosity function for objects in the CMD
diagram, using a bin width of $0.12$\,mag. {\bf Upper panel:} colour
distribution for objects in the CMD. The grey curve represents the
smoothed distribution, and the dashed vertical lines indicate the
colour peaks for blue and red GC candidates, respectively, for comparison 
purposes.}    
\label{dcol2}    
\end{figure}

In Figure\,\ref{espa3} is shown the projected spatial distribution for
candidates brighter than $i'_0= 24$\,mag and with typical colours  of 
GCs (see Section\,\ref{candsel}). The plot presents the usual orientation,
North is up, East to the left. The grey scale indicates the smooth
projected density, and the symbols correspond to objects bluer than 
$(g'-i')_0=0.9$ (blue circles), redder than $(g'-i')_0=1$ (red squares), 
and presenting intermediate colours (green diamonds). The selection of these
limits is based on the colour distribution shown in Figure\,\ref{dcol2}.
 
The spatial distribution is densely concentrated towards the galaxy centre,
with few objects at projected distances larger than 90\,arcsec from the
galaxy centre (dashed circle). The area contained by the dashed circle 
represents $\approx22$ per cent of the FOV. As a comparison, the objects 
brighter than $i'_0= 24$\,mag enclosed in this area correspond to $\approx76$ 
per cent of the total, while the blue GC candidates fainter than 
$i'_0= 24$\,mag achieve the $51$ per cent and the red CG candidates in the 
same magnitude range represent the $61$ per cent. From these results, a 
proportion test points to statistically different results with probabilities
of $p=99$ and $p=97$ per cent, respectively. 
A Kolmogorov-Smirnov test \citep{kol33} indicates that the distributions 
differ with probabilities of $p=98$ per cent and $p=91$ per cent, respectively.
Taking into account that we are dealing with small sample statistic, 
we might subdivide the objects 
brighter than $i'_0= 24$\,mag into the three colour ranges previously indicated.
The bluer group is poorly populated and its members are not so strongly 
concentrated towards the galaxy centre ($\approx 55\%$) in comparison with 
those presenting intermediate and red colours ($\approx 82\%$).

The bottom left panel in Figure\,\ref{dcol2} shows the CMD for those 
objects brighter than 
$i'_0= 24$\,mag with galactocentric distances up to 90\,arcsec. It 
presents a clear gap around $i'_0 = 23.4$\,mag, also evident in the 
luminosity function at the right panel. The objects above and below the
gap fulfill similar colour ranges, with slightly different mean colours.
Objects fainter than $i'_0 = 23.4$\,mag present a mean $(g'-i')_0$ 
colour of $\approx 1$\,mag, while the mean colour for the brighter ones 
is $\approx 1.05$\,mag, with comparable dispersion in both cases 
($\sigma \approx 0.14$ and $\sigma \approx 0.16$, respectively). 
Objects brighter than 
$i'_0 = 23.4$\,mag have a mean galactocentric distance of $\approx 
60$\,arcsec and all of them except one are located at distances from the 
galaxy centre larger than 30\,arcsec (highlighted with framed symbols 
in Figure\,\ref{espa3}). On the other hand, the mean galactocentric 
distance for the fainter ones is $\approx 40$\,arcsec and around one 
third are located at less than 30\,arcsec from the galaxy centre.

The upper panel shows the colour distribution for objects in the CMD
with a bin size of 0.04\,mag, and the smoothed distribution with solid 
lines. The dashed vertical lines indicate the mean colour for the
GCs in the bimodal case from Table\,\ref{fit}. A significant
fraction of the sample presents intermediate colours, with a narrow
distribution in $(g'-i')_0 = 0.9 - 1$\,mag. The comparison of this
colour range with the expected colours from \citet{bre12} SSP models, 
with a \citet{cha01}
lognormal IMF, points to younger SSPs or presenting a lower metal content
than the results from \citet{gon15}. For a SSP of $\approx 3.5$\,Gyr the
colour range corresponds to $0.40 \lesssim Z/Z_{\odot} \lesssim 0.75$,
while a metallicity of $Z/Z_{\odot} \approx 1.9$ corresponds to SSPs
from $\approx 2$\,Gyr to $\approx 2.6$\,Gyr.

Answering the question whether there is an excess of bright objects, 
we calculate the expected number of GCs for the corresponding magnitude 
range. From 
the set of parameters fitted to the GCLF in Section\,\ref{lfsec},
the fraction of GCs brighter than $i'_0=24$\,mag is $0.03\pm0.007$.
From the numerical integration of the radial profile for circular
symmetry derived in Section\,\ref{radsec}, the GC candidates at 
lower galactocentric distances than 90\,arcsec represent the 
$\approx 55\%$ of the total population. Hence, the expected number
of GCs brighter than $i'_0= 24$\,mag up to that radial limit should
be $\approx 12\pm1$ GCs. In order to consider possible changes 
in the GCLF due the relaxed limits in $\chi^2$, we calculated again 
the completeness curves and the GCLF for candidates fainter than
$i'_0= 24$\,mag. The fitted TOM and width, 
$i'_{0,TOM}=25.39\pm0.08\pm0.1$ and $\sigma=0.77\pm0.08\pm0.1$,
did not vary from those calculated in Section\,\ref{lfsec} when
uncertainties are considered. In this case, the expected number
of GCs brighter than $i'_0= 24$\,mag up to galactocentric 
distances of 90\,arcsec was $\approx 14\pm5$ GCs. In both cases
the expected number were significantly lower than the 41 objects
found in the photometry. In a different approach, we generated 
1000 random samples of GC candidates from a Gaussian profile 
with the parameters previously fitted to the GCLF. In only $\approx 3\%$ 
of the cases were obtained more than 20 objects brighter than 
$i'_0= 24$\,mag, without samples with 40 objects brighter 
than $i'_0= 24$\,mag. If we consider the variation of the 
Gaussian parameters according to their uncertainties, in just $1\%$ of
the cases the probability of obtaining 40 objects brighter 
than $i'_0= 24$\,mag was larger than $5\%$.

Hence, this analysis supports the existence of an excess of bright GC
candidates, in comparison with the expected numbers obtained from the
GCLF of the fainter ones.

\section{Discussion}

Unlike other findings for many iEs \citep[e.g.][]{tal09}, 
NGC\,6411 lacks photometric evidence for a recent merger event  
like tidal features or striking changes in the surface profile 
parameters. Its surface brightness distribution is well represented by undisturbed
ellipses ($A4 \approx 0$), unlike the usual scenario in recent merger 
remnants \citep[e.g.][]{bas17}. From integration to infinity of the
$g'$ and $i'$ galaxy luminosity profiles, we obtained $(g'-i')_0\approx0.9$
and, if we just consider the inner 30\,arcsec, $(g'-i')_0\approx0.99$. 
These colours are in agreement
with those expected from the population derived by \citep{gon15}, whose age 
and metallicity correspond to $(g'-i')_0\approx1.1$ according to the SSP
models from CMD\,3.0 \citep{bre12} and the MILES\footnote{http://miles.iac.es/pages/ssp-models.php} stellar population models \citep{vaz10}. 

Considering the population parameters derived by \citet{gon15},
\citet{vaz10} population models lead to $M/L_V\approx 1.82$, which 
implies for NGC\,6411 a stellar mass of ${\rm M_{\star} \approx 7 \times 
10^{10}\,M_{\odot}}$. According to the colour profile, the stellar 
component is expected to be dominated by a population somewhat older, and 
the stellar mass should range from this value to few times 
$10^{11}\,{\rm M_{\odot}}$.

This galaxy also presents a supernova, SN199da, classified by
\citet{fil99} as subluminous SNIa. This SNe subtype might be 
associated with early-type hosts and old stellar populations 
\citep{how01}. Moreover, \citet{gal08}
indicate a correlation with the host galaxy age and metallicity, 
where galaxies with higher iron abundance host less luminous SNe,
while SN in galaxies with a characteristic age greater than 
$\approx5$\,Gyr are $\approx1$\,mag fainter at their maximum, 
a lower limit that matches with the age derived by \citet{san06}
for NGC\,6411.

Its GCS is poorly populated in comparison with Es of similar
luminosity in groups/clusters \citep[e.g][]{har13,cas17}, in agreement 
with other ellipticals in low density environments \citep[e.g.][]{cas13a,
sal15,bas17}. The $S_N$ for iEs with similar luminosity 
than NGC\,6411 typically ranges from 1 to 1.5 \citep{fos11,sal15},
while cluster and group Es span up to larger values of $S_N$
\citep[][and references there in]{for98,pen08,har13}.
 The fraction of metal-rich GCs in NGC\,6411 represents 
$\approx 40\%$ of the population, typical of bright Es 
\citep{har15}, even in low density environments \citep{sal15}. 

The extreme environmental conditions needed in the build-up of GCs point
to the relevance of galaxy mergers in their formation and 
survival in the early stages \citep{kru14,kru15}. This implies a direct 
connection between a GCS and the evolutionary history of its host galaxy.
The current paradigm for GC formation states that the metal-poor GCs
in early-type galaxies formed in major star-formation episodes during 
the merger of the first proto-galaxies, or were accreted from satellite 
galaxies.
On the other hand, the origin of the metal-rich GCs occurred in
later stages of the galaxy evolution, during a few massive mergers, 
involving more evolved galaxies \citep{mur10,ton13,li14}. 
In this sense, \citet{pen06} found that the colour range spanned by a 
GCS correlates with the galaxy stellar mass.

The mean colours corresponding to GC
subpopulations also seem to correlate with galaxy stellar mass, 
particularly the red ones \citep[e.g.][]{str04a,fai11}. 
According with the correlation of the mean colours of GC populations 
with their host galaxy luminosity from \citet{fai11}, the expected 
values for the blue and red GC populations in NGC\,6411 are $(g'-i')_0 
\approx 0.78$ and $1.07$, respectively. These only differ few hundredths 
from the colour peaks obtained in Section\,\ref{colsec}, pointing
that the majority of the GCs candidates fainter than $i'_0=24$\,mag fit to
typical colours for old GCs.
Despite of this, the distance between the peak colours for blue and 
red GCs in NGC\,6411 is slightly smaller, corresponding
to fainter galaxies.

The TOM of the GCLF is in agreement with SBF measurements 
\citep{bla01}, which also points to an old GCS. 
The width of the GCLF also corresponds to a galaxy with lower
stellar mass, according to the relations derived by \citet{har14}.
This is in agreement with the assumption that NGC\,6411 has suffered
an intermediate age merger, 
with a corresponding stellar population contributing to 
its luminosity. The small dispersion in the GCLF might also provide 
clues about the evolutionary history of NGC\,6411, pointing to minor 
mergers as main contributors to the GCS, particularly to the metal-poor
subpopulation. In addition, wet mergers, would 
have supplied gas to form the metal-rich GCs. If we assume a power-law 
as the initial mass function of GCs \citep[e.g.][]{broc14}, the amount 
of gas available as a consequence of minor mergers
might result in marginal probabilities of forming massive GCs. Moreover, 
the low-density environment spread the occurrence of subsequent mergers,
hence reduces the efficacy of mergers to redistribute young GCs and
facilitate their survival \citep{kru15}. The disruption affects
more severely low-mass GCs, turning the power-law IMF into a bell
shaped distribution \citep{ros16}. Hence, the absence of successive mergers
might lead to a larger fraction of disruption for the low-mass GCs,
which translates in a lower dispersion of the GCLF. At the high-mass
end, GCs might result from the merger of star clusters complex like
those observed in some merger remnants \citep[e.g.][]{whi99,fel05},
but it is not well established how efficient is this process to form 
bright GCs.

We can also compare the GCS extension with the properties of the host 
galaxy applying the relations from \citet{kar14} for early-type
galaxies. A GCS with a projected extension of $\approx 70$\,kpc,
corresponds to a galaxy with a ${\rm R_{eff}}\approx 6$\,kpc 
(equation\,11) and a stellar mass of 
${\rm M_{\star} \approx 5\times10^{11}\,M_{\odot}}$ (equation\,10). From 
the surface brightness profiles in the three filters we obtained 
${\rm R_{eff}}\approx 6.4$\,kpc, which is in agreement with the expected
value. Regarding its stellar mass, as we previously indicated in Section\,
\ref{lfsec} the uncertain ages and metallicities for the galaxy population
result in a wide range of stellar masses. However, the assumption that the 
stellar mass is dominated by an older and less metal-rich population than 
that proposed by \citet{gon15} for the inner region seems plausible according 
with the galaxy overall characteristics.
Other issue to comment is the marginal difference in the slopes of radial
distributions in blue and red GCs. Observational studies of rich GCSs in 
bright ellipticals show a metal-rich subpopulation more concentrated towards 
the galaxy, while the metal-poor one presents a flattened distribution 
\citep[e.g.][]{dir05,bas06a,fos11}. This behaviour is also reproduced in 
numerical models \citep[e.g.][]{amo18}. Results might differ for 
non-standard ellipticals
and lenticulars with evidence of multiple populations \citep[e.g.][]{cas13b,
cas15a,ses16}. These galaxies presented evidence of intermediate age 
populations due to late mergers, wich might have reconfigured the distribution
of typical GC subpopulations. However, we notice that this characteristic
is not common in all the merger remnants \citep[e.g.][]{bas17}, and further
analysis should be carried on.

The bright GC candidates present in NGC\,6411 point to a late merger,
in agreement with results from \citet{gon15} and \citet{san06}. As
previously indicated, the intermediate colours are fulfilled by a wide 
range of SSPs, but do not simultaneously match with the ages and
metallicities proposed by \citet{gon15}. The metal content would imply
a young population, $\approx 2$\,Gyr. The age of $\approx 3.5$\,Gyr 
corresponds to greater metallicities than the usual values for old metal-rich 
GCs, denoting an enrichment of the gas, but lower than expected for bright 
ellipticals. This is in agreement with the bluer colour of NGC\,6411 in
comparison with ellipticals with similar stellar mass \citep{lac16}.
Regarding the absolute magnitudes of these bright GCs, assuming the 
distance modulus derived in Section\,\ref{lfsec}, they span from $M_{i'}=-9.3$
to $M_{i'}=-10.7$\,mag. Despite these objects are fainter than UCDs
in other elliptical galaxies, they fulfil the luminosity range of UCDs 
in \citet{bro11}, which were selected from their effective radii instead 
of luminosity criteria. If we use equations\,1 and 2 from \citet{bas17} to 
estimate $(V,I)$ magnitudes, the resulting absolute magnitudes range 
from $M_V=-8.6$ to $M_V=-10$\,mag, fainter than $\omega$\,Cen \citep[][2010 Edition]{har96}. 
This difference could be larger when intermediate ages are assumed for 
bright GCs in NGC\,6411, as can be realized from the CMD\,3.0 SSP models. 
For instance, a population of 3.5\,Gyr and $Z=0.7\,Z_{\odot}$ presents
$(g'-i')=0.98$\,mag and $i'=4.93$\,mag, getting $\Delta_{i'}=0.92$\,mag
fainter when aged to 10\,Gyr. For a lower metallicity, $Z=0.5\,Z_{\odot}$,
an age around $4-4.5$\,Gyr results in $(g'-i')=0.95-0.96$\,mag, fading 
$\Delta_{i'}=0.8-0.7$\,mag when aged to 10\,Gyr.
Hence, they would be brighter than the TOM magnitude for
old GCs, but their luminosities would not achieve those of typical UCDs. 
A wet merger at intermediate-age might
be responsible for the existence of the bright GCs candidates. However,
this event should have occured in earlier stages than the age derived by
\citet{gon15} for the
stellar population in the inner region of NGC\,6411, driving the starburst 
that formed the brightest objects.

\section{Summary} 

On the basis of $g'$, $r'$ and $i'$ photometry obtained with GEMINI GMOS-N,
we carried out a study of NGC\,6411 and its GCS down to the turn-over 
magnitude. We summarize our conclusions in the following.

\begin{itemize}
\item 
The galaxy does not present striking evidence of merger events, 
no tidal structures were found after subtracting a smoothed 
surface brightness component, neither in the galaxy colour map. 
There is no sign of boxy isophotes according to the $A4$ parameter. 

\item Despite not examining the full spatial extent of the GCS, 
our observations do allow us to determine the projected density radial 
profile, and the GCS size. Results are in agreement with literature 
compilations for galaxies with similar luminosity. The total population 
is estimated to be $720\pm40$ GCs. The $S_N$ points to a poor GCS, as 
usually found in low density environments.

\item The fraction of red GCs does not differ from those found in brighter
ellipticals with larger GCS, but the difference between the colour peaks
and the dispersion of the GCLF are slightly lower. We interpret this as
evidence of the relevance of minor mergers in the build up of the galaxy.

\item The TOM of the GCLF is in agreement with SBF studies. The colour 
distribution resembles those found in typical bright Es. These
findings characterise an old GCS, which is in agreement with the 
absence of recent merger evidence.

\item We detect an excess of bright GCs, strongly concentrated
towards the galaxy. Their colours are mainly intermediate between the
peaks for the blue and red population. Their colours are slightly bluer 
than the galaxy in the inner region, we interpret that their 
origin might be related to a past merger. Their colours do not agree
with ages derived by the spectroscopic studies of the diffuse
light of the galaxy, pointing to an older event.
\end{itemize}

\section*{Acknowledgments}
This research was funded with grants from Consejo Nacional de 
Investigaciones Cient\'{\i}ficas y T\'ecnicas de la Rep\'ublica 
Argentina (PIP 112-201101-00393), Agencia Nacional de Promoci\'on 
Cient\'{\i}fica y Tecnol\'ogica (PICT-2013-0317), and Universidad 
Nacional de La Plata (UNLP 11-G150), Argentina. TR acknowledges 
support from CONICYT project Basal AFB-170002.

\bibliographystyle{mnras}
\bibliography{biblio}

\begin{thebibliography}{}
\makeatletter
\relax
\def\mn@urlcharsother{\let\do\@makeother \do\$\do\&\do\#\do\^\do\_\do\%\do\~}
\def\mn@doi{\begingroup\mn@urlcharsother \@ifnextchar [ {\mn@doi@}
  {\mn@doi@[]}}
\def\mn@doi@[#1]#2{\def\@tempa{#1}\ifx\@tempa\@empty \href
  {http://dx.doi.org/#2} {doi:#2}\else \href {http://dx.doi.org/#2} {#1}\fi
  \endgroup}
\def\mn@eprint#1#2{\mn@eprint@#1:#2::\@nil}
\def\mn@eprint@arXiv#1{\href {http://arxiv.org/abs/#1} {{\tt arXiv:#1}}}
\def\mn@eprint@dblp#1{\href {http://dblp.uni-trier.de/rec/bibtex/#1.xml}
  {dblp:#1}}
\def\mn@eprint@#1:#2:#3:#4\@nil{\def\@tempa {#1}\def\@tempb {#2}\def\@tempc
  {#3}\ifx \@tempc \@empty \let \@tempc \@tempb \let \@tempb \@tempa \fi \ifx
  \@tempb \@empty \def\@tempb {arXiv}\fi \@ifundefined
  {mn@eprint@\@tempb}{\@tempb:\@tempc}{\expandafter \expandafter \csname
  mn@eprint@\@tempb\endcsname \expandafter{\@tempc}}}

\bibitem[\protect\citeauthoryear{{Amorisco}}{{Amorisco}}{2018}]{amo18}
{Amorisco} N.~C.,  2018, preprint, \href
  {http://adsabs.harvard.edu/abs/2018arXiv180200812A} {} (\mn@eprint {arXiv}
  {1802.00812})

\bibitem[\protect\citeauthoryear{{Bamford} et~al.,}{{Bamford}
  et~al.}{2009}]{bam09}
{Bamford} S.~P.,  et~al., 2009, \mn@doi [\mnras]
  {10.1111/j.1365-2966.2008.14252.x}, \href
  {http://adsabs.harvard.edu/abs/2009MNRAS.393.1324B} {393, 1324}

\bibitem[\protect\citeauthoryear{{Bassino} \& {Caso}}{{Bassino} \&
  {Caso}}{2017}]{bas17}
{Bassino} L.~P.,  {Caso} J.~P.,  2017, \mn@doi [MNRAS] {10.1093/mnras/stw3390},
  \href {http://adsabs.harvard.edu/abs/2017MNRAS.466.4259B} {466, 4259}

\bibitem[\protect\citeauthoryear{{Bassino}, {Faifer}, {Forte}, {Dirsch},
  {Richtler}, {Geisler}  \& {Schuberth}}{{Bassino} et~al.}{2006}]{bas06a}
{Bassino} L.~P.,  {Faifer} F.~R.,  {Forte} J.~C.,  {Dirsch} B.,  {Richtler} T.,
   {Geisler} D.,   {Schuberth} Y.,  2006, \mn@doi [A\&A]
  {10.1051/0004-6361:20054563}, \href
  {http://adsabs.harvard.edu/abs/2006A%26A...451..789B} {451, 789}

\bibitem[\protect\citeauthoryear{{Bertin} \& {Arnouts}}{{Bertin} \&
  {Arnouts}}{1996}]{ber96}
{Bertin} E.,  {Arnouts} S.,  1996, A\&AS, \href
  {http://adsabs.harvard.edu/abs/1996A%26AS..117..393B} {117, 393}

\bibitem[\protect\citeauthoryear{{Binney} \& {Tremaine}}{{Binney} \&
  {Tremaine}}{1987}]{bin87}
{Binney} J.,  {Tremaine} S.,  1987, {Galactic dynamics}

\bibitem[\protect\citeauthoryear{{Blakeslee}, {Lucey}, {Barris}, {Hudson}  \&
  {Tonry}}{{Blakeslee} et~al.}{2001}]{bla01}
{Blakeslee} J.~P.,  {Lucey} J.~R.,  {Barris} B.~J.,  {Hudson} M.~J.,   {Tonry}
  J.~L.,  2001, \mn@doi [MNRAS] {10.1046/j.1365-8711.2001.04800.x}, \href
  {http://adsabs.harvard.edu/abs/2001MNRAS.327.1004B} {327, 1004}

\bibitem[\protect\citeauthoryear{{Bressan}, {Marigo}, {Girardi}, {Salasnich},
  {Dal Cero}, {Rubele}  \& {Nanni}}{{Bressan} et~al.}{2012}]{bre12}
{Bressan} A.,  {Marigo} P.,  {Girardi} L.,  {Salasnich} B.,  {Dal Cero} C.,
  {Rubele} S.,   {Nanni} A.,  2012, \mn@doi [MNRAS]
  {10.1111/j.1365-2966.2012.21948.x}, \href
  {http://adsabs.harvard.edu/abs/2012MNRAS.427..127B} {427, 127}

\bibitem[\protect\citeauthoryear{{Brockamp}, {K{\"u}pper}, {Thies}, {Baumgardt}
   \& {Kroupa}}{{Brockamp} et~al.}{2014}]{broc14}
{Brockamp} M.,  {K{\"u}pper} A.~H.~W.,  {Thies} I.,  {Baumgardt} H.,   {Kroupa}
  P.,  2014, \mn@doi [MNRAS] {10.1093/mnras/stu562}, \href
  {http://adsabs.harvard.edu/abs/2014MNRAS.441..150B} {441, 150}

\bibitem[\protect\citeauthoryear{{Brodie}, {Romanowsky}, {Strader}  \&
  {Forbes}}{{Brodie} et~al.}{2011}]{bro11}
{Brodie} J.~P.,  {Romanowsky} A.~J.,  {Strader} J.,   {Forbes} D.~A.,  2011,
  \mn@doi [AJ] {10.1088/0004-6256/142/6/199}, \href
  {http://adsabs.harvard.edu/abs/2011AJ....142..199B} {142, 199}

\bibitem[\protect\citeauthoryear{{Br{\"u}ns} \& {Kroupa}}{{Br{\"u}ns} \&
  {Kroupa}}{2012}]{bru12}
{Br{\"u}ns} R.~C.,  {Kroupa} P.,  2012, \mn@doi [A\&A]
  {10.1051/0004-6361/201219693}, \href
  {http://adsabs.harvard.edu/abs/2012A%26A...547A..65B} {547, A65}

\bibitem[\protect\citeauthoryear{{Caso}, {Bassino}, {Richtler}, {Smith
  Castelli}  \& {Faifer}}{{Caso} et~al.}{2013a}]{cas13a}
{Caso} J.~P.,  {Bassino} L.~P.,  {Richtler} T.,  {Smith Castelli} A.~V.,
  {Faifer} F.~R.,  2013a, \mn@doi [MNRAS] {10.1093/mnras/sts687}, \href
  {http://adsabs.harvard.edu/abs/2013MNRAS.430.1088C} {430, 1088}

\bibitem[\protect\citeauthoryear{{Caso}, {Richtler}, {Bassino}, {Salinas},
  {Lane}  \& {Romanowsky}}{{Caso} et~al.}{2013b}]{cas13b}
{Caso} J.~P.,  {Richtler} T.,  {Bassino} L.~P.,  {Salinas} R.,  {Lane} R.~R.,
  {Romanowsky} A.,  2013b, \mn@doi [A\&A] {10.1051/0004-6361/201321032}, \href
  {http://adsabs.harvard.edu/abs/2013A%26A...555A..56C} {555, A56}

\bibitem[\protect\citeauthoryear{{Caso}, {Bassino}  \& {G{\'o}mez}}{{Caso}
  et~al.}{2015}]{cas15a}
{Caso} J.~P.,  {Bassino} L.~P.,   {G{\'o}mez} M.,  2015, \mn@doi [MNRAS]
  {10.1093/mnras/stv2015}, \href
  {http://adsabs.harvard.edu/abs/2015MNRAS.453.4421C} {453, 4421}

\bibitem[\protect\citeauthoryear{{Caso}, {Bassino}  \& {G{\'o}mez}}{{Caso}
  et~al.}{2017}]{cas17}
{Caso} J.~P.,  {Bassino} L.~P.,   {G{\'o}mez} M.,  2017, \mn@doi [\mnras]
  {10.1093/mnras/stx1393}, \href
  {http://adsabs.harvard.edu/abs/2017MNRAS.470.3227C} {470, 3227}

\bibitem[\protect\citeauthoryear{{Chabrier}}{{Chabrier}}{2001}]{cha01}
{Chabrier} G.,  2001, \mn@doi [ApJ] {10.1086/321401}, \href
  {http://adsabs.harvard.edu/abs/2001ApJ...554.1274C} {554, 1274}

\bibitem[\protect\citeauthoryear{{Ciotti}}{{Ciotti}}{1991}]{cio91}
{Ciotti} L.,  1991, A\&A, \href
  {http://adsabs.harvard.edu/abs/1991A%26A...249...99C} {249, 99}

\bibitem[\protect\citeauthoryear{{Dirsch}, {Richtler}  \& {Bassino}}{{Dirsch}
  et~al.}{2003}]{dir03b}
{Dirsch} B.,  {Richtler} T.,   {Bassino} L.~P.,  2003, \mn@doi [A\&A]
  {10.1051/0004-6361:20031027}, \href
  {http://adsabs.harvard.edu/abs/2003A%26A...408..929D} {408, 929}

\bibitem[\protect\citeauthoryear{{Dirsch}, {Schuberth}  \& {Richtler}}{{Dirsch}
  et~al.}{2005}]{dir05}
{Dirsch} B.,  {Schuberth} Y.,   {Richtler} T.,  2005, \mn@doi [A\&A]
  {10.1051/0004-6361:20035737}, \href
  {http://adsabs.harvard.edu/abs/2005A%26A...433...43D} {433, 43}

\bibitem[\protect\citeauthoryear{{Dressler}}{{Dressler}}{1980}]{dre80}
{Dressler} A.,  1980, \mn@doi [\apj] {10.1086/157753}, \href
  {http://adsabs.harvard.edu/abs/1980ApJ...236..351D} {236, 351}

\bibitem[\protect\citeauthoryear{{Escudero}, {Faifer}, {Bassino}, {Calder\'on}
  \& {Caso}}{{Escudero} et~al.}{2015}]{esc15}
{Escudero} C.~G.,  {Faifer} F.~R.,  {Bassino} L.~P.,  {Calder\'on} J.~P.,
  {Caso} J.~P.,  2015, MNRAS, 0, 0

\bibitem[\protect\citeauthoryear{{Faifer} et~al.,}{{Faifer}
  et~al.}{2011}]{fai11}
{Faifer} F.~R.,  et~al., 2011, \mn@doi [MNRAS]
  {10.1111/j.1365-2966.2011.19018.x}, \href
  {http://adsabs.harvard.edu/abs/2011MNRAS.416..155F} {416, 155}

\bibitem[\protect\citeauthoryear{{Fellhauer} \& {Kroupa}}{{Fellhauer} \&
  {Kroupa}}{2005}]{fel05}
{Fellhauer} M.,  {Kroupa} P.,  2005, \mn@doi [MNRAS]
  {10.1111/j.1365-2966.2005.08891.x}, \href
  {http://adsabs.harvard.edu/abs/2005MNRAS.359..223F} {359, 223}

\bibitem[\protect\citeauthoryear{{Filippenko}}{{Filippenko}}{1999}]{fil99}
{Filippenko} A.~V.,  1999, IAU Circ, \href
  {http://adsabs.harvard.edu/abs/1999IAUC.7219....1F} {7219}

\bibitem[\protect\citeauthoryear{{Forbes}, {Grillmair}, {Williger}, {Elson}  \&
  {Brodie}}{{Forbes} et~al.}{1998}]{for98}
{Forbes} D.~A.,  {Grillmair} C.~J.,  {Williger} G.~M.,  {Elson} R.~A.~W.,
  {Brodie} J.~P.,  1998, \mn@doi [MNRAS] {10.1046/j.1365-8711.1998.01202.x},
  \href {http://adsabs.harvard.edu/abs/1998MNRAS.293..325F} {293, 325}

\bibitem[\protect\citeauthoryear{{Foster} et~al.,}{{Foster}
  et~al.}{2011}]{fos11}
{Foster} C.,  et~al., 2011, \mn@doi [MNRAS] {10.1111/j.1365-2966.2011.18965.x},
  \href {http://adsabs.harvard.edu/abs/2011MNRAS.415.3393F} {415, 3393}

\bibitem[\protect\citeauthoryear{{Gallagher}, {Garnavich}, {Caldwell},
  {Kirshner}, {Jha}, {Li}, {Ganeshalingam}  \& {Filippenko}}{{Gallagher}
  et~al.}{2008}]{gal08}
{Gallagher} J.~S.,  {Garnavich} P.~M.,  {Caldwell} N.,  {Kirshner} R.~P.,
  {Jha} S.~W.,  {Li} W.,  {Ganeshalingam} M.,   {Filippenko} A.~V.,  2008,
  \mn@doi [ApJ] {10.1086/590659}, \href
  {http://adsabs.harvard.edu/abs/2008ApJ...685..752G} {685, 752}

\bibitem[\protect\citeauthoryear{{Gonz{\'a}lez Delgado} et~al.,}{{Gonz{\'a}lez
  Delgado} et~al.}{2015}]{gon15}
{Gonz{\'a}lez Delgado} R.~M.,  et~al., 2015, \mn@doi [A\&A]
  {10.1051/0004-6361/201525938}, \href
  {http://adsabs.harvard.edu/abs/2015A%26A...581A.103G} {581, A103}

\bibitem[\protect\citeauthoryear{{Harris}}{{Harris}}{1996}]{har96}
{Harris} W.~E.,  1996, \mn@doi [AJ] {10.1086/118116}, \href
  {http://adsabs.harvard.edu/abs/1996AJ....112.1487H} {112, 1487}

\bibitem[\protect\citeauthoryear{{Harris} \& {van den Bergh}}{{Harris} \& {van
  den Bergh}}{1981}]{har81}
{Harris} W.~E.,  {van den Bergh} S.,  1981, \mn@doi [AJ] {10.1086/113047},
  \href {http://adsabs.harvard.edu/abs/1981AJ.....86.1627H} {86, 1627}

\bibitem[\protect\citeauthoryear{{Harris}, {Kavelaars}, {Hanes}, {Pritchet}  \&
  {Baum}}{{Harris} et~al.}{2009}]{har09c}
{Harris} W.~E.,  {Kavelaars} J.~J.,  {Hanes} D.~A.,  {Pritchet} C.~J.,   {Baum}
  W.~A.,  2009, \mn@doi [\aj] {10.1088/0004-6256/137/2/3314}, \href
  {http://adsabs.harvard.edu/abs/2009AJ....137.3314H} {137, 3314}

\bibitem[\protect\citeauthoryear{{Harris}, {Harris}  \& {Alessi}}{{Harris}
  et~al.}{2013}]{har13}
{Harris} W.~E.,  {Harris} G.~L.~H.,   {Alessi} M.,  2013, \mn@doi [ApJ]
  {10.1088/0004-637X/772/2/82}, \href
  {http://adsabs.harvard.edu/abs/2013ApJ...772...82H} {772, 82}

\bibitem[\protect\citeauthoryear{{Harris} et~al.,}{{Harris}
  et~al.}{2014}]{har14}
{Harris} W.~E.,  et~al., 2014, \mn@doi [ApJ] {10.1088/0004-637X/797/2/128},
  \href {http://adsabs.harvard.edu/abs/2014ApJ...797..128H} {797, 128}

\bibitem[\protect\citeauthoryear{Harris, Harris  \& Hudson}{Harris
  et~al.}{2015}]{har15}
Harris W.~E.,  Harris G.~L.,   Hudson M.~J.,  2015, \mn@doi [The Astrophysical
  Journal] {10.1088/0004-637X/806/1/36}, 806, 36

\bibitem[\protect\citeauthoryear{{Hern{\'a}ndez-Toledo}, {V{\'a}zquez-Mata},
  {Mart{\'{\i}}nez-V{\'a}zquez}, {Avila Reese}, {M{\'e}ndez-Hern{\'a}ndez},
  {Ortega-Esbr{\'{\i}}}  \& {N{\'u}{\~n}ez}}{{Hern{\'a}ndez-Toledo}
  et~al.}{2008}]{her08}
{Hern{\'a}ndez-Toledo} H.~M.,  {V{\'a}zquez-Mata} J.~A.,
  {Mart{\'{\i}}nez-V{\'a}zquez} L.~A.,  {Avila Reese} V.,
  {M{\'e}ndez-Hern{\'a}ndez} H.,  {Ortega-Esbr{\'{\i}}} S.,   {N{\'u}{\~n}ez}
  J.~P.~M.,  2008, \mn@doi [\aj] {10.1088/0004-6256/136/5/2115}, \href
  {http://adsabs.harvard.edu/abs/2008AJ....136.2115H} {136, 2115}

\bibitem[\protect\citeauthoryear{{Hirschmann}, {De Lucia}, {Iovino}  \&
  {Cucciati}}{{Hirschmann} et~al.}{2013}]{hir13}
{Hirschmann} M.,  {De Lucia} G.,  {Iovino} A.,   {Cucciati} O.,  2013, \mn@doi
  [\mnras] {10.1093/mnras/stt827}, \href
  {http://adsabs.harvard.edu/abs/2013MNRAS.433.1479H} {433, 1479}

\bibitem[\protect\citeauthoryear{{Howell}}{{Howell}}{2001}]{how01}
{Howell} D.~A.,  2001, \mn@doi [ApJL] {10.1086/321702}, \href
  {http://adsabs.harvard.edu/abs/2001ApJ...554L.193H} {554, L193}

\bibitem[\protect\citeauthoryear{{Jim{\'e}nez}, {Cora}, {Bassino}, {Tecce}  \&
  {Smith Castelli}}{{Jim{\'e}nez} et~al.}{2011}]{jim11}
{Jim{\'e}nez} N.,  {Cora} S.~A.,  {Bassino} L.~P.,  {Tecce} T.~E.,   {Smith
  Castelli} A.~V.,  2011, \mn@doi [MNRAS] {10.1111/j.1365-2966.2011.19328.x},
  \href {http://adsabs.harvard.edu/abs/2011MNRAS.417..785J} {417, 785}

\bibitem[\protect\citeauthoryear{{Jord{\'a}n} et~al.,}{{Jord{\'a}n}
  et~al.}{2007}]{jor07}
{Jord{\'a}n} A.,  et~al., 2007, \mn@doi [ApJs] {10.1086/516840}, \href
  {http://adsabs.harvard.edu/abs/2007ApJS..171..101J} {171, 101}

\bibitem[\protect\citeauthoryear{{Kartha}, {Forbes}, {Spitler}, {Romanowsky},
  {Arnold}  \& {Brodie}}{{Kartha} et~al.}{2014}]{kar14}
{Kartha} S.~S.,  {Forbes} D.~A.,  {Spitler} L.~R.,  {Romanowsky} A.~J.,
  {Arnold} J.~A.,   {Brodie} J.~P.,  2014, \mn@doi [MNRAS]
  {10.1093/mnras/stt1880}, \href
  {http://adsabs.harvard.edu/abs/2014MNRAS.437..273K} {437, 273}

\bibitem[\protect\citeauthoryear{{Kolmogorov}}{{Kolmogorov}}{1933}]{kol33}
{Kolmogorov} A.,  1933, {Giornale dell'Istituto Italiano degli Attuari}, 4, 83

\bibitem[\protect\citeauthoryear{{Kruijssen}}{{Kruijssen}}{2014}]{kru14}
{Kruijssen} J.~M.~D.,  2014, \mn@doi [Classical and Quantum Gravity]
  {10.1088/0264-9381/31/24/244006}, \href
  {http://adsabs.harvard.edu/abs/2014CQGra..31x4006K} {31, 244006}

\bibitem[\protect\citeauthoryear{{Kruijssen}}{{Kruijssen}}{2015}]{kru15}
{Kruijssen} J.~M.~D.,  2015, \mn@doi [MNRAS] {10.1093/mnras/stv2026}, \href
  {http://adsabs.harvard.edu/abs/2015MNRAS.454.1658K} {454, 1658}

\bibitem[\protect\citeauthoryear{{Lacerna}, {Hern{\'a}ndez-Toledo},
  {Avila-Reese}, {Abonza-Sane}  \& {del Olmo}}{{Lacerna} et~al.}{2016}]{lac16}
{Lacerna} I.,  {Hern{\'a}ndez-Toledo} H.~M.,  {Avila-Reese} V.,  {Abonza-Sane}
  J.,   {del Olmo} A.,  2016, \mn@doi [\aap] {10.1051/0004-6361/201527844},
  \href {http://adsabs.harvard.edu/abs/2016A%26A...588A..79L} {588, A79}

\bibitem[\protect\citeauthoryear{{Lane}, {Salinas}  \& {Richtler}}{{Lane}
  et~al.}{2013}]{lan13}
{Lane} R.~R.,  {Salinas} R.,   {Richtler} T.,  2013, \mn@doi [A\&A]
  {10.1051/0004-6361/201220231}, \href
  {http://adsabs.harvard.edu/abs/2013A%26A...549A.148L} {549, A148}

\bibitem[\protect\citeauthoryear{{Li} \& {Gnedin}}{{Li} \&
  {Gnedin}}{2014}]{li14}
{Li} H.,  {Gnedin} O.~Y.,  2014, \mn@doi [ApJ] {10.1088/0004-637X/796/1/10},
  \href {http://adsabs.harvard.edu/abs/2014ApJ...796...10L} {796, 10}

\bibitem[\protect\citeauthoryear{{Marcum}, {Aars}  \& {Fanelli}}{{Marcum}
  et~al.}{2004}]{marc04}
{Marcum} P.~M.,  {Aars} C.~E.,   {Fanelli} M.~N.,  2004, \mn@doi [\aj]
  {10.1086/420708}, \href {http://adsabs.harvard.edu/abs/2004AJ....127.3213M}
  {127, 3213}

\bibitem[\protect\citeauthoryear{{M{\'e}ndez}, {Teodorescu}, {Kudritzki}  \&
  {Burkert}}{{M{\'e}ndez} et~al.}{2009}]{men09}
{M{\'e}ndez} R.~H.,  {Teodorescu} A.~M.,  {Kudritzki} R.-P.,   {Burkert} A.,
  2009, \mn@doi [\apj] {10.1088/0004-637X/691/1/228}, \href
  {http://adsabs.harvard.edu/abs/2009ApJ...691..228M} {691, 228}

\bibitem[\protect\citeauthoryear{{Muratov} \& {Gnedin}}{{Muratov} \&
  {Gnedin}}{2010}]{mur10}
{Muratov} A.~L.,  {Gnedin} O.~Y.,  2010, \mn@doi [ApJ]
  {10.1088/0004-637X/718/2/1266}, \href
  {http://adsabs.harvard.edu/abs/2010ApJ...718.1266M} {718, 1266}

\bibitem[\protect\citeauthoryear{{Niemi}, {Hein{\"a}m{\"a}ki}, {Nurmi}  \&
  {Saar}}{{Niemi} et~al.}{2010}]{nie10}
{Niemi} S.-M.,  {Hein{\"a}m{\"a}ki} P.,  {Nurmi} P.,   {Saar} E.,  2010,
  \mn@doi [MNRAS] {10.1111/j.1365-2966.2010.16457.x}, \href
  {http://adsabs.harvard.edu/abs/2010MNRAS.405..477N} {405, 477}

\bibitem[\protect\citeauthoryear{{Norris} et~al.,}{{Norris}
  et~al.}{2014}]{nor14}
{Norris} M.~A.,  et~al., 2014, \mn@doi [MNRAS] {10.1093/mnras/stu1186}, \href
  {http://adsabs.harvard.edu/abs/2014MNRAS.443.1151N} {443, 1151}

\bibitem[\protect\citeauthoryear{{Peng} et~al.,}{{Peng} et~al.}{2006}]{pen06}
{Peng} E.~W.,  et~al., 2006, \mn@doi [ApJ] {10.1086/498210}, \href
  {http://adsabs.harvard.edu/abs/2006ApJ...639...95P} {639, 95}

\bibitem[\protect\citeauthoryear{{Peng} et~al.,}{{Peng} et~al.}{2008}]{pen08}
{Peng} E.~W.,  et~al., 2008, \mn@doi [ApJ] {10.1086/587951}, \href
  {http://adsabs.harvard.edu/abs/2008ApJ...681..197P} {681, 197}

\bibitem[\protect\citeauthoryear{{Richtler}}{{Richtler}}{2003}]{ric03}
{Richtler} T.,  2003, in {Alloin} D.,  {Gieren} W.,  eds,  Lecture Notes in
  Physics, Berlin Springer Verlag Vol. 635, Stellar Candles for the
  Extragalactic Distance Scale. pp 281--305 (\mn@eprint {}
  {arXiv:astro-ph/0304318}), \mn@doi{10.1007/978-3-540-39882-0_15}

\bibitem[\protect\citeauthoryear{{Richtler}}{{Richtler}}{2013}]{ric13}
{Richtler} T.,  2013, in {Pugliese} G.,  {de Koter} A.,   {Wijburg} M.,  eds,
  Astronomical Society of the Pacific Conference Series Vol. 470, 370 Years of
  Astronomy in Utrecht. p.~327 (\mn@eprint {arXiv} {1210.0045})

\bibitem[\protect\citeauthoryear{{Richtler}, {Bassino}, {Dirsch}  \&
  {Kumar}}{{Richtler} et~al.}{2012}]{ric12a}
{Richtler} T.,  {Bassino} L.~P.,  {Dirsch} B.,   {Kumar} B.,  2012, \mn@doi
  [A\&A] {10.1051/0004-6361/201118589}, \href
  {http://adsabs.harvard.edu/abs/2012A%26A...543A.131R} {543, A131}

\bibitem[\protect\citeauthoryear{{Richtler}, {Hilker}, {Kumar}, {Bassino},
  {G{\'o}mez}  \& {Dirsch}}{{Richtler} et~al.}{2014}]{ric14}
{Richtler} T.,  {Hilker} M.,  {Kumar} B.,  {Bassino} L.~P.,  {G{\'o}mez} M.,
  {Dirsch} B.,  2014, \mn@doi [A\&A] {10.1051/0004-6361/201423525}, \href
  {http://adsabs.harvard.edu/abs/2014A%26A...569A..41R} {569, A41}

\bibitem[\protect\citeauthoryear{{Richtler}, {Salinas}, {Lane}, {Hilker}  \&
  {Schirmer}}{{Richtler} et~al.}{2015}]{ric15}
{Richtler} T.,  {Salinas} R.,  {Lane} R.~R.,  {Hilker} M.,   {Schirmer} M.,
  2015, \mn@doi [A\&A] {10.1051/0004-6361/201424530}, \href
  {http://adsabs.harvard.edu/abs/2015A%26A...574A..21R} {574, A21}

\bibitem[\protect\citeauthoryear{{Robin}, {Reyl{\'e}}, {Derri{\`e}re}  \&
  {Picaud}}{{Robin} et~al.}{2003}]{rob03}
{Robin} A.~C.,  {Reyl{\'e}} C.,  {Derri{\`e}re} S.,   {Picaud} S.,  2003,
  \mn@doi [\aap] {10.1051/0004-6361:20031117}, \href
  {http://cdsads.u-strasbg.fr/abs/2003A%26A...409..523R} {409, 523}

\bibitem[\protect\citeauthoryear{{Rossi}, {Bekki}  \& {Hurley}}{{Rossi}
  et~al.}{2016}]{ros16}
{Rossi} L.~J.,  {Bekki} K.,   {Hurley} J.~R.,  2016, \mn@doi [\mnras]
  {10.1093/mnras/stw1827}, \href
  {http://adsabs.harvard.edu/abs/2016MNRAS.462.2861R} {462, 2861}

\bibitem[\protect\citeauthoryear{{Salinas}, {Richtler}, {Bassino}, {Romanowsky}
   \& {Schuberth}}{{Salinas} et~al.}{2012}]{sal12}
{Salinas} R.,  {Richtler} T.,  {Bassino} L.~P.,  {Romanowsky} A.~J.,
  {Schuberth} Y.,  2012, \mn@doi [A\&A] {10.1051/0004-6361/201116517}, \href
  {http://adsabs.harvard.edu/abs/2012A%26A...538A..87S} {538, A87}

\bibitem[\protect\citeauthoryear{{Salinas}, {Alabi}, {Richtler}  \&
  {Lane}}{{Salinas} et~al.}{2015}]{sal15}
{Salinas} R.,  {Alabi} A.,  {Richtler} T.,   {Lane} R.~R.,  2015, \mn@doi
  [A\&A] {10.1051/0004-6361/201425574}, \href
  {http://adsabs.harvard.edu/abs/2015A%26A...577A..59S} {577, A59}

\bibitem[\protect\citeauthoryear{{S{\'a}nchez-Bl{\'a}zquez}
  et~al.,}{{S{\'a}nchez-Bl{\'a}zquez} et~al.}{2006}]{san06}
{S{\'a}nchez-Bl{\'a}zquez} P.,  et~al., 2006, \mn@doi [MNRAS]
  {10.1111/j.1365-2966.2006.10699.x}, \href
  {http://adsabs.harvard.edu/abs/2006MNRAS.371..703S} {371, 703}

\bibitem[\protect\citeauthoryear{{Schawinski} et~al.,}{{Schawinski}
  et~al.}{2014}]{sch14}
{Schawinski} K.,  et~al., 2014, \mn@doi [\mnras] {10.1093/mnras/stu327}, \href
  {http://adsabs.harvard.edu/abs/2014MNRAS.440..889S} {440, 889}

\bibitem[\protect\citeauthoryear{{Schlafly} \& {Finkbeiner}}{{Schlafly} \&
  {Finkbeiner}}{2011}]{sch11}
{Schlafly} E.~F.,  {Finkbeiner} D.~P.,  2011, \mn@doi [ApJ]
  {10.1088/0004-637X/737/2/103}, \href
  {http://adsabs.harvard.edu/abs/2011ApJ...737..103S} {737, 103}

\bibitem[\protect\citeauthoryear{{Sersic}}{{Sersic}}{1968}]{ser68}
{Sersic} J.~L.,  1968, {Atlas de galaxias australes}

\bibitem[\protect\citeauthoryear{{Sesto}, {Faifer}  \& {Forte}}{{Sesto}
  et~al.}{2016}]{ses16}
{Sesto} L.~A.,  {Faifer} F.~R.,   {Forte} J.~C.,  2016, \mn@doi [\mnras]
  {10.1093/mnras/stw1627}, \href
  {http://adsabs.harvard.edu/abs/2016MNRAS.461.4260S} {461, 4260}

\bibitem[\protect\citeauthoryear{{Smith} \& et al.}{{Smith} \&
  et~al.}{2002}]{smi02}
{Smith} J.~A.,  et al. 2002, \mn@doi [AJ] {10.1086/339311}, \href
  {http://adsabs.harvard.edu/abs/2002AJ....123.2121S} {123, 2121}

\bibitem[\protect\citeauthoryear{{Spitler}, {Forbes}, {Strader}, {Brodie}  \&
  {Gallagher}}{{Spitler} et~al.}{2008}]{spi08}
{Spitler} L.~R.,  {Forbes} D.~A.,  {Strader} J.,  {Brodie} J.~P.,   {Gallagher}
  J.~S.,  2008, \mn@doi [MNRAS] {10.1111/j.1365-2966.2007.12823.x}, \href
  {http://adsabs.harvard.edu/abs/2008MNRAS.385..361S} {385, 361}

\bibitem[\protect\citeauthoryear{{Stetson}}{{Stetson}}{1987}]{ste87}
{Stetson} P.~B.,  1987, \mn@doi [PASP] {10.1086/131977}, \href
  {http://adsabs.harvard.edu/abs/1987PASP...99..191S} {99, 191}

\bibitem[\protect\citeauthoryear{{Strader}, {Brodie}  \& {Forbes}}{{Strader}
  et~al.}{2004}]{str04a}
{Strader} J.,  {Brodie} J.~P.,   {Forbes} D.~A.,  2004, \mn@doi [AJ]
  {10.1086/420995}, \href {http://adsabs.harvard.edu/abs/2004AJ....127.3431S}
  {127, 3431}

\bibitem[\protect\citeauthoryear{{Tal}, {van Dokkum}, {Nelan}  \&
  {Bezanson}}{{Tal} et~al.}{2009}]{tal09}
{Tal} T.,  {van Dokkum} P.~G.,  {Nelan} J.,   {Bezanson} R.,  2009, \mn@doi
  [AJ] {10.1088/0004-6256/138/5/1417}, \href
  {http://adsabs.harvard.edu/abs/2009AJ....138.1417T} {138, 1417}

\bibitem[\protect\citeauthoryear{{Tonini}}{{Tonini}}{2013}]{ton13}
{Tonini} C.,  2013, \mn@doi [ApJ] {10.1088/0004-637X/762/1/39}, \href
  {http://adsabs.harvard.edu/abs/2013ApJ...762...39T} {762, 39}

\bibitem[\protect\citeauthoryear{{Tully} et~al.,}{{Tully} et~al.}{2013}]{tul13}
{Tully} R.~B.,  et~al., 2013, \mn@doi [AJ] {10.1088/0004-6256/146/4/86}, \href
  {http://adsabs.harvard.edu/abs/2013AJ....146...86T} {146, 86}

\bibitem[\protect\citeauthoryear{{Vazdekis}, {S{\'a}nchez-Bl{\'a}zquez},
  {Falc{\'o}n-Barroso}, {Cenarro}, {Beasley}, {Cardiel}, {Gorgas}  \&
  {Peletier}}{{Vazdekis} et~al.}{2010}]{vaz10}
{Vazdekis} A.,  {S{\'a}nchez-Bl{\'a}zquez} P.,  {Falc{\'o}n-Barroso} J.,
  {Cenarro} A.~J.,  {Beasley} M.~A.,  {Cardiel} N.,  {Gorgas} J.,   {Peletier}
  R.~F.,  2010, \mn@doi [MNRAS] {10.1111/j.1365-2966.2010.16407.x}, \href
  {http://adsabs.harvard.edu/abs/2010MNRAS.404.1639V} {404, 1639}

\bibitem[\protect\citeauthoryear{{Whitmore}, {Zhang}, {Leitherer}, {Fall},
  {Schweizer}  \& {Miller}}{{Whitmore} et~al.}{1999}]{whi99}
{Whitmore} B.~C.,  {Zhang} Q.,  {Leitherer} C.,  {Fall} S.~M.,  {Schweizer} F.,
    {Miller} B.~W.,  1999, \mn@doi [AJ] {10.1086/301041}, \href
  {http://adsabs.harvard.edu/abs/1999AJ....118.1551W} {118, 1551}

\bibitem[\protect\citeauthoryear{{de Vaucouleurs}, {de Vaucouleurs}, {Corwin},
  {Buta}, {Paturel}  \& {Fouqu{\'e}}}{{de Vaucouleurs} et~al.}{1991}]{dev91}
{de Vaucouleurs} G.,  {de Vaucouleurs} A.,  {Corwin} Jr. H.~G.,  {Buta} R.~J.,
  {Paturel} G.,   {Fouqu{\'e}} P.,  1991, {Third Reference Catalogue of Bright
  Galaxies. Volume I: Explanations and references. Volume II: Data for galaxies
  between 0$^{h}$ and 12$^{h}$. Volume III: Data for galaxies between 12$^{h}$
  and 24$^{h}$.}

\bibitem[\protect\citeauthoryear{{van Dokkum} et~al.,}{{van Dokkum}
  et~al.}{2010}]{vdo10}
{van Dokkum} P.~G.,  et~al., 2010, \mn@doi [ApJ]
  {10.1088/0004-637X/709/2/1018}, \href
  {http://adsabs.harvard.edu/abs/2010ApJ...709.1018V} {709, 1018}

\makeatother
\end{thebibliography}

\label{lastpage}
\end{document}